\documentclass[11pt,fleqn]{article} 
\usepackage{graphicx}
\usepackage{textcomp} 
\usepackage{epstopdf}
\usepackage{latexsym}
\usepackage{amsmath}
\usepackage{amssymb}
\usepackage{bm}
\usepackage[mathcal]{euscript}
\usepackage{slashed}
\allowdisplaybreaks

\usepackage[top=30mm, bottom=30mm, left=25mm, right=25mm]{geometry}

\numberwithin{equation}{section}

\usepackage{marvosym}

\usepackage{indentfirst}

\newcommand\blfootnote[1]{
  \begingroup
  \renewcommand\thefootnote{}\footnote{#1}
  \addtocounter{footnote}{-1}
  \endgroup
}

\usepackage{hyperref}
\hypersetup{
        unicode,	
        colorlinks,
        citecolor=blue,
        linkcolor=blue, 
        urlcolor=red,
        bookmarksopen=true,
        bookmarksopenlevel=\maxdimen,
      }

\def\gl#1#2{\ifmmode \mathrm{GL}(#1; {\bf #2}) \else $\mathrm{GL}(#1; {\bf #2})$\fi}
\def\sl#1#2{\ifmmode \mathrm{SL}(#1; {\bf #2}) \else $\mathrm{SL}(#1; {\bf #2})$\fi}
\def\so#1{\ifmmode \mathrm{SO}({#1}) \else $\mathrm{SO}(#1)$\fi}

\def\sp#1#2{\ifmmode \mathrm{Sp}(#1; {\bf #2}) \else $\mathrm{Sp}(#1; {\bf #2})$\fi}
\def\usp#1{\ifmmode \mathrm{USp}(#1) \else $\mathrm{USp}(#1)$\fi}
\def\spin#1{\ifmmode \mathrm{Spin}(#1) \else $\mathrm{Spin}(#1)$\fi}
\def\su#1{\ifmmode \mathrm{SU}({#1}) \else $\mathrm{SU}(#1)$\fi}

\def\double #1{#1{\hbox{\kern-2pt $#1$}}}

\def\ket#1{\left| #1\right>}

\def\dd{\hbox{\,\Large$\triangleright$}}

\mathcode`\*="702A                  
\def\half{{\textstyle{1\over{\raise.1ex\hbox{$\scriptstyle{2}$}}}}}

\def \p{\partial}
\def \a{\alpha}
\def \b{\beta}
\def \e{\epsilon}
\def \d{\delta}

\def \g{\gamma}
\def \l{\lambda}
\def \s{\sigma}
\def \lb{\bar\lambda}
\def \L{\Lambda}

\def \o{\omega}
\def \ob{\bar\omega}
\def \O{\Omega}
\def \t{\theta}

\def \Zb{{\mathbb Z}}
\def \psu{{\mathfrak{psu}}}
\def \frakh{{\mathfrak{h}}}
\def \frakg{{\mathfrak{g}}}
\def \frakl{{\mathfrak{l}}}

\def \sT{{\mathsf T}}
\def \sM{{\mathsf M}}
\def \sP{{\mathsf P}}
\def \sQ{{\mathsf Q}}
\def \sJ{{\mathsf J}}
\def \sE{{\mathsf E}}
\def \sS{{\mathsf S}}
\def \sA{{\mathsf A}}
\def \sD{{\mathsf D}}
\def \sq{{\mathsf q}}
\def \ss{{\mathsf s}}
\def \sp{{\mathsf p}}
\def \se{{\mathsf e}}
\def \st{{\mathsf t}}
\def \tsq{{\tilde\sq}}
\def\ad{{\dot a}}
\def\bd{{\dot b}}
\def\cd{{\dot c}}
\def\dd{{\dot d}}
\def\im{{\rm i}}
\def\ua{{\underline a}}
\def\ub{{\underline b}}
\def\uc{{\underline c}}
\def\ud{{\underline d}}

\def\uab{{\underline ab}}

\begin{document}

\begin{flushright}
\makebox[0pt][b]{}
\end{flushright}

\vspace{40pt}
\begin{center}
{\LARGE Vertex operators for the plane wave pure spinor string}

\vspace{40pt}
Osvaldo Chandia${}^\clubsuit$ and Brenno Carlini Vallilo${}^{\spadesuit}$
\vspace{40pt}

{\em ${}^\clubsuit$ Departamento de Ciencias, Facultad de Artes Liberales \& UAI Physics Center,\\ Universidad Adolfo Ib\'a\~nez, Chile
\vspace{10pt}
\\${}^{\spadesuit}$ Departamento de Ciencias F\'{\i}sicas, Universidad Andres Bello, \\
Sazie 2212, Santiago, Chile
}\\

\vspace{60pt}
{\bf Abstract}
\end{center}
In this work we give an explicit construction for the vertex operators
of massless states in the pure spinor superstring in a plane wave background. The construction is based on the observation that the full action can be divided in two parts, where the simpler one is based on a smaller coset and closely resembles the gauge fixed Green-Schwarz action.

\blfootnote{\\
${}^\clubsuit$ \href{mailto:ochandiaq@gmail.com}{ochandiaq@gmail.com}\\
${}^{\spadesuit}$ \href{mailto:vallilo@gmail.com}{vallilo@gmail.com} }

\setcounter{page}0
\thispagestyle{empty}

\newpage

\tableofcontents

\parskip = 0.1in
\section{Introduction}

For more than ten years there has been enormous progress in the
understanding of both sides of the AdS/CFT conjecture due to the
presence of integrable structures
\cite{Beisert:2010jr,Arutyunov:2009ga}.
Even after all the progress it is still not known
how to obtain the physical spectrum and amplitudes at finite
$AdS$ radius.

In principle one could use perturbation theory at large radius using
the pure spinor description for the superstring. Some partial results
have been obtained for the massless spectrum
\cite{Bedoya:2010qz,Mikhailov:2011af,Berkovits:2012ps}, but a complete
dictionary between BPS states and the corresponding vertex operator
is still missing. An attempt to describe a massive state was made in
\cite{Vallilo:2011fj}, however some contributions were incorrectly
ignored \cite{Heinze:2015xxa}. Although the pure spinor
sigma model is classically integrable
\cite{Vallilo:2003nx,Chandia:2016ueo} and some quantum aspects have
been studied \cite{Mikhailov:2007eg,Benichou:2011ch} there are no
techniques available to help computing the  spectrum. It is possible
that the  formalism
developed in \cite{Ashok:2009xx} and applied in
\cite{Benichou:2010rk,Eberhardt:2018exh} could be generalized
to the case of $AdS$ pure spinor string.

Before attacking the $AdS$ case we could first look at the simpler
 BMN limit \cite{Berenstein:2002jq}. The
Green-Schwarz superstring in this space was studied extensively
\cite{Metsaev:2001bj,Metsaev:2002re} and pure spinor string was
studied in \cite{Berkovits:2002zv} and \cite{Chandia:2014wca}. The
approach of \cite{Chandia:2014wca} was to use the background
field expansion of the usual $AdS_5\times S^5$ pure spinor string
around the BPS state with $\sE=\sJ$. Although the resulting model
was free and the full spectrum can be computed, the isometries of
the plane wave background were not manifest in this approach and the
BMN limit spontaneously breaks conformal invariance. The latter
severely reduces its usefulness.

In this paper we will study vertex operators for the string defined in
\cite{Berkovits:2002zv}. Although the sigma model is not as simple as
the free gauge fixed Green-Schwarz version, its structure simplifies
some computations. For example, it was proved in
\cite{Berkovits:2002zv} that the beta function has only a vanishing
one loop contribution. We will find an explicit construction for the
unintegrated vertex operator using the isometries of the
background. We then compute the integrated vertex operator using the
standard BRST descent procedure. Although the final expression is not
very illuminating, the plane wave background admits D9-branes and the
open string version is simple and could lead to
the construction of DDF-like operators \cite{Jusinskas:2014vqa}.

This paper is organized as follows. In Section \ref{review} we review
how to obtain the isometry algebra of the plane wave limit of
$AdS_5\times S^5$ background as a contraction of the $\psu(2,2|4)$
algebra. In Section \ref{sec:supergeo} we describe the supergeometry
of the plane wave background and compute the covariant derivatives and
symmetry generators. The pure spinor description of the superstring in
BMN limit of $AdS_5\times S^5$ is reviewed in Section \ref{sugra}. In
Section~\ref{sec:Masslessvertex} we construct the unintegrated vertex
operator for all BPS states using part of the BRST charge and the
isometries of the background. Finally in Section \ref{sec:IVO} we use
the standard descent procedure to find the general form of the
integrated vertex operator. The appendices contain conventions
and some details left out from the body of the paper.

\section{BMN limit of the $\psu(2,2|4)$ algebra}
\label{review}

The isometry algebra of the BMN limit \cite{Berenstein:2002jq} of
$AdS_5\times S^5$ is obtained as a contraction of the $\psu(2,2|4)$
algebra \cite{Hatsuda:2002kx}, which can be understood
geometrically as the Penrose limit of
the original $AdS$ space \cite{Blau:2001ne,Blau:2002dy}.
The idea is to look
for the isometries a massless particle with very high energy
in $AdS_5$ and  very high angular momentum in $S^5$ sees. Let us
first look at the bosonic subalgebra generated by
$(\sM_{AB},\sM_A,\sP_A,\sT,\sM_{IJ},\sM_I,\sP_I,\sJ)$ which
is $\mathfrak{so}(2,4)\oplus\mathfrak{so}(6)$ in an
$\mathfrak{so}(4)\oplus \mathfrak{so}(4)$ basis. Their commutators
can be found in the Appendix \ref{app1}. We chose a convention where
the translations $(\sP_A,\sT,\sP_I,\sJ)$ are hermitian and the
rotations $( \sM_{AB},\sM_A,\sM_{IJ},\sM_I)$ are anti-hermitian.
Note that in the limit $R\to\infty$ the algebra defined by
these generators is $\mathfrak{iso}(1,4)\oplus\mathfrak{iso}(5)$.
This is expected to be enhanced to the full $\mathfrak{iso}(1,9)$
since new conserved currents will appear as $R\to\infty$.

Since the particle is massless the eigenvalues of $\sE$ and $\sJ$ should be the same. So in order to have a generator the survives this limit we define
\begin{align}
  \sE_+ = R(\sT + \sJ).
\end{align}
Furthermore, since their sum will diverge we define
\begin{align}
         \sE_- =R^{-1}(\sT-\sJ),
\end{align}
where $R$ is the radius of $AdS_5$ and $S^5$. Both of these generators will be well defined in the limit $R\to\infty$. Now we invert these definitions
\begin{align}
  \sT=\frac{1}{2}\left(R\sE_- +R^{-1}\sE_+\right),\quad
  \sJ=\frac{1}{2} \left(R^{-1}\sE_+-R\sE_-\right).
\end{align}
Boosts in the $\sT$ and $\sJ$ directions $\sM_A$ and $\sM_I$ also have to be re-scaled
\begin{align}
      \sM_A =- \im R\bar\sP_A,\quad \sM_I= \im R \bar\sP_I,
\end{align}
now $\bar\sP_A$ and $\bar\sP_I$ are hermitian. The reason for this
notation will become clear later; we will be able to organize the
generators in representations of $\mathfrak{su}(2|2)\oplus\mathfrak{su}(2|2)$.

Using these definitions and the commutators in Appendix \ref{app1}, after taking the $R\to\infty$ limit, the non-vanishing commutators are
(besides the $\mathfrak{so}(4)\oplus \mathfrak{so}(4)$ algebra generated by $\sM_{AB}$ and $\sM_{IJ}$, which is left unchanged):
\begin{align}
&[\sP_A,\bar\sP_B]=-\frac{\im}{2}\delta_{AB}\sE_-,\quad
[\sP_I,\bar\sP_J]=-\frac{\im}{2}\delta_{IJ}\sE_-\\
&[\sE_+,\sP_A]=\im\bar\sP_A,\quad [\sE_+,\bar\sP_A]=-\im\sP_A,\\
&[\sE_+,\sP_I]=\im\bar\sP_I,\quad [\sE_+,\bar\sP_I]=-\im\sP_I.
\end{align}

We can see that the generators
$(\sP_A,\bar\sP_A,\sP_I,\bar\sP_I,\sE_-)$ form a Heisenberg algebra
$\mathfrak{h}(8)$ with central element $\sE_-$ and $\sE_+$ acts as an
outer automorphism of the the algebra that commutes with
$\mathfrak{so}(4)\oplus \mathfrak{so}(4)$. It should be stressed that
the $\mathfrak{so}(4)\oplus \mathfrak{so}(4)$ algebra is {\em not}
promoted to full $\mathfrak{so}(8)$ in the limit $R\to\infty$.
The reason for this is the presence of the supercharges, which
we will now describe.

We will first define new scalings for the supercharges
$(\sQ_a,\sQ_\ad,\bar\sQ_a,\bar\sQ_\ad)$. We will try
\begin{align}\label{newScaleQ}
\sS_a = R^{\frac12}\sQ_a.\quad \bar\sS_a=R^{\frac12} \bar\sQ_a,\quad
\sS_\ad = R^{-\frac12}\sQ_\ad,\quad
\bar\sS_\ad = R^{-\frac12}\bar\sQ_\ad.
\end{align}
With these definitions we obtain the expected supersymmetry algebra
\begin{align}
 & \{\sS_a,\sS_b\}= \{\bar\sS_a,\bar\sS_b\} =\delta_{ab}\sE_+,\qquad\qquad
   \{\sS_\ad,\sS_\bd\}= \{\bar\sS_\ad,\bar\sS_\bd\}
  =\delta_{\ad\bd}\sE_-,\\
 & \{\sS_a,\sS_\ad\}=\{\bar\sS_a,\bar\sS_\ad\}= \sigma^A_{a\ad}\sP_A+\sigma^I_{a\ad}\sP_I,
\end{align}
with $\sE_-$ still playing the role of a central charge. Also note
that  $(\sS_a,\sS_\ad,\bar\sS_a,\bar\sS_\ad)$ are still
$\mathfrak{so}(4)\oplus\mathfrak{so}(4)$ spinors and have the expected commutators with
the generators $(\sM_{AB},\sM_{IJ})$. Furthermore, with (\ref{newScaleQ}) and taking the
$R\to\infty$ limit we obtain
\begin{align}
  &[\sE_+, \sS_a]=[\sE_+, \bar\sS_a]=0,\\
  & [\sP_A,\sS_\ad]=[\sP_A,\bar\sS_\ad]=
    [\bar\sP_A,\sS_\ad]=[\bar\sP_A,\bar\sS_\ad]=0,\\
  & [\sP_I,\sS_\ad]=[\sP_I,\bar\sS_\ad]=
    [\bar\sP_I,\sS_\ad]=[\bar\sP_I,\bar\sS_\ad]=0,\\
   &[\sE_+, \sS_\ad]=\im\Pi_{\ad\bd}\bar\sS_\bd,\quad
    [\sE_+, \bar\sS_\ad]=-\im\Pi_{\ad\bd}\sS_\bd,\\
  & [\sP_A,\sS_a]=-\frac{\im}{2} (\sigma_A\Pi)_{a\ad}\bar\sS_\ad, \quad
    [\sP_A,\bar\sS_a]=\frac{\im}{2} (\sigma_A\Pi)_{a\ad}\sS_\ad,\\
  & [\sP_I,\sS_a]=-\frac{\im}{2} (\sigma_I\Pi)_{a\ad}\bar\sS_\ad, \quad
    [\sP_I,\bar\sS_a]=\frac{\im}{2} (\sigma_I\Pi)_{a\ad}\sS_\ad,\\
  & [\bar\sP_A,\sS_a]=\frac{\im}{2} (\sigma_A)_{a\ad}\sS_\ad, \quad
    [\bar\sP_A,\bar\sS_a]=\frac{\im}{2} (\sigma_A)_{a\ad}\bar\sS_\ad,\\
 & [\bar\sP_I,\sS_a]=\frac{\im}{2} (\sigma_I)_{a\ad}\sS_\ad, \quad
    [\bar\sP_I,\bar\sS_a]=\frac{\im}{2} (\sigma_I)_{a\ad}\bar\sS_\ad,
\end{align}
and finally we have that
\begin{align}
  &\{\sS_\ad,\bar\sS_\bd\}=0,\\
  &\{ \sS_a , \bar\sS_b \}  = -\frac{\im}{2} \left( (\s^{AB}\Pi)_{ab}\sM_{AB} - (\s^{IJ}\Pi)_{ab}  \sM_{IJ} \right) ,\\
  &\{ \sS_a , \bar\sS_\bd \} = (\s^A\Pi)_{a\bd} \bar\sP_A
  + (\s^I\Pi)_{a\bd} \bar\sP_I  ,\\
  &\{ \sS_\ad , \bar\sS_a \} = -(\s^A\Pi)_{a
    \ad}\bar\sP_A
    - (\s^I\Pi)_{a \ad} \bar\sP_I,
\end{align}
where $(\sigma^i\Pi)$ always means $(\sigma^i)_{a\bd}\Pi_{\bd\ad}$ and
$\Pi$ is symmetric, traceless and squares to identity.
This concludes the contraction of the $\psu(2,2|4)$ algebra.

\subsection{Organizing in terms of
  $\mathfrak{psu}(2|2)\oplus\mathfrak{psu}(2|2)\oplus\mathbb{R}^2$}
\label{sec:su22su22}

We can see that $\sE_+$ acts as a rotation operator for
$(\sP_A,\sP_I,\sS_\ad)\times
(\bar\sP_A,\bar\sP_I,\bar\sS_\ad)$. Furthermore, the generators
$(\sM_{AB},\sM_{IJ})$ also rotates this set. It turns out that the we
can organize all generators in a Jordan structure
\cite{Gunaydin:1984fk}
$\frakl^-\oplus\frakl^0\oplus\frakl^+$ that satisfies
\footnote{$\frakl^0$ should not be confused with $\frakg_0$
in the $\mathbb{Z}_4$ decomposition.}
\begin{align}\label{Jordan}
  &[\frakl^0,\frakl^0]\subset\frakl^0,\quad
  [\frakl^0,\frakl^-]\subset\frakl^-,\quad
  [\frakl^0,\frakl^+]\subset\frakl^+,\quad
  [\frakl^+,\frakl^-]\subset\frakl^0,\cr
  &[\frakl^+,\frakl^+]=[\frakl^-,\frakl^-]=\emptyset
\end{align}
where $\frakl^0$ is
$\psu(2|2)\oplus\psu(2|2)\oplus\mathfrak{u}(1)\oplus\mathbb{R}$ and is
generated by $\{\sM_{AB},\sM_{IJ},\sS_a,\bar\sS_a,\sE_+,\sE_-\}$. We are
using that $\mathfrak{so}(4)\oplus\mathfrak{so}(4)\simeq
\mathfrak{su}(2)\oplus\mathfrak{su}(2)
\oplus\mathfrak{su}(2)\oplus\mathfrak{su}(2)$. Before describing $\frakl^+$ and $\frakl^-$ we have to define the following combinations
\begin{align}
  &\sA_A = \sP_A -\im\bar\sP_A,\quad \sA_A^\dagger =\sP_A
  +\im\bar\sP_A,\quad \sA_I=\sP_I -\im\bar\sP_I,\quad
  \sA_I^\dagger=\sP_I +\im\bar\sP_I,\\
  &\sD_\ad=\frac{1}{\sqrt{2}}\left(\sS_\ad -\im\Pi_{\ad\bd}\bar\sS_\bd\right),\quad
  \sD_\ad^\dagger=\frac{1}{\sqrt{2}}\left(\sS_\ad +\im\Pi_{\ad\bd}\bar\sS_\bd\right)
\end{align}
that satisfy a super Heisenberg algebra $\frakh(8|8)$ with central
element $\sE_-$. The full Jordan decomposition is
\begin{align}
&\frakl^0=\psu(2|2)\oplus\psu(2|2)\oplus\mathfrak{u}(1)
\oplus\mathbb{R} =
\{\sM_{AB},\sM_{IJ},\sS_a,\bar\sS_a,\sE_+,\sE_-\},\\
&\frakl^+ =\{\sA^\dagger_A,\sA^\dagger_J,\sD^\dagger_\ad \}, \quad
\frakl^- =\{\sA_A,\sA_J,\sD_\ad \}.
\end{align}
It is straightforward to check that the relations (\ref{Jordan}) are satisfied.

Another interesting property of the contraction is that there exist a
closed sub-algebra that has the same $\Zb_4$ decomposition of the
original $\psu(2,2|4)$. The decomposition is
\begin{align}
  \frakg_0=\{\bar\sP_A,\bar\sP_I\},\quad \frakg_1=\{\sS_\ad\},\quad
  \frakg_2=\{\sE_+,\sE_-,\sP_A,\sP_I\},\quad  \frakg_3=\{\bar\sS_\ad\}.
\end{align}

The algebra generated by the operators above is
$\frak{h}(8|8)\rtimes\mathfrak{u}(1)$ where the $\mathfrak{u}(1)$ is generated
by $\sE_+$. Will see the the coset
$(H(8|8)\rtimes U(1))/
(\mathbb{R}^4\times\mathbb{R}^4)$, where
$\mathbb{R}^4\times\mathbb{R}^4$ is generated by
$\{\bar\sP_A,\bar\sP_I\}$, plays an important role.

\subsection{Casimir of the contracted algebra and spectrum}
\label{sec:casimirSpec}

The quadratic Casimir for the $\psu(2,2|4)$ algebra written with the original generators is
\begin{align}
  \mathfrak{C}^{\psu}_2=& -\sT^2 + \delta^{AB}\sP_A\sP_B + \sJ^2+
   \delta^{IJ}\sP_I\sP_J -\frac{1}{R^2} \delta^{AB}\sM_A\sM_B -
   \frac{1}{R^2}\delta^{IJ}\sM_I\sM_J +\cr
  &-\frac{1}{2R^2} \delta^{AC}\delta^{BD}\sM_{AB}\sM_{CD}+
  \frac{1}{2R^2}\delta^{IK}\delta^{JL}\sM_{IJ}\sM_{KL}
    -\frac{\im}{R} \Pi^{ab}\sQ_a\bar\sQ_b-
    \frac{\im}{R} \Pi^{\ad\bd}\sQ_\ad\bar\sQ_\bd.
\end{align}
In the limit $R\to\infty$ we get the quadratic Casimir of the $d=10$ super Poincar\'e algebra $-\sT^2 + \delta^{AB}\sP_A\sP_B + \sJ^2+
  \delta^{IJ}\sP_I\sP_J$. If we use the re-scalings defined before, we see that the surviving terms are
\begin{align}\label{casimirH88}
  \mathfrak{C}_2= -\sE_+\sE_- + \delta^{AB}\sP_A\sP_B +
  \delta^{IJ}\sP_I\sP_J + \delta^{AB}\bar\sP_A\bar\sP_B +
  \delta^{IJ}\bar\sP_I\bar\sP_J  - \im\Pi^{\ad\bd}\sS_\ad\bar\sS_\bd.
\end{align}
One can check that is commutes with all generators. Note that the
Casimir operator only contains generators of
$\frak{h}(8|8)\rtimes\mathfrak{u}(1)$. This indicates that all important physics happens in the smaller coset $(H(8|8)\rtimes U(1))/
(\mathbb{R}^4\times\mathbb{R}^4)$. This is also related to the fact that the variables $(\theta^a,\bar\theta^a)$ that would be gauge fixed using kappa symmetry in the GS string.

Using the definitions of $\frakl^+$ and $\frakl^-$ above
 we can write the Casimir as
\begin{align}
  \mathfrak{C}_2= -\sE_-\sE_+ + \delta^{AB}\sA_A^\dagger\sA_B +
  \delta^{IJ}\sA_I^\dagger\sA_J+ \delta^{\ad\bd}\sD_\ad^\dagger\sD_\bd.
\end{align}
which has a more familiar form. We can build representations of the algebra starting with a vacuum $\ket{E_-}$ satisfying \cite{Gunaydin:1984fk}
\begin{align}
  \sE_-\ket{E_-}=E_-\ket{E_-},\quad
  \sA_A\ket{E_-}=\sA_I\ket{E_-}=\sD_\ad\ket{E_-}=0.
\end{align}
Note that contrary to the flat space superstring the vacuum is not
degenerate since $\sD_\ad^\dagger$ changes the value of the energy
measured in spacetime. The Casimir operator, which should correspond
to the sum of the zero modes of the Virasoro operators $L_0+\bar L_0$
\cite{Chandia:2014wca},
kills $\ket{E_-}$ if the eigenvalue of $\sE_+$ is equal to $0$. We can
change this acting on $\ket{E_-}$ with $\sD_\ad^\dagger$. This will change
the eigenvalue of $\sE^-$ from $0$ to $8$. The multiplet obtained in
this way is the supergravity multiplet \cite{Metsaev:2002re}
which is composed of
$(8+8)\times(8+8)=1+8+28+56+70+56+28+8+1=256$ states.
More explicitly, using that
\begin{align}
  [\delta^{\ad\bd}\sD^\dagger_\ad\sD_\bd,\sD^\dagger_\cd]=-\sE_-\sD^\dagger_\cd,\quad
  [\sE_+,\sD^\dagger_\ad]= \sD_\ad^\dagger,\quad
  [\mathfrak{C}_2, \sD^\dagger_\ad]=0,
\end{align}
the excited states are given by
\begin{align}
  \ket{E_-,\ad}= \sD^\dagger_\ad\ket{E_-},\quad
  \ket{E_-,\ad\bd}= \sD^\dagger_\bd\sD^\dagger_\ad\ket{E_-},\quad
  \ket{E_-,\ad\bd\cd}=\sD_\cd^\dagger
  \sD^\dagger_\bd\sD^\dagger_\ad\ket{E_-},\quad {\rm etc.}
\end{align}
and they all satisfy
\begin{align}
  \mathfrak{C}_2 \ket{\Psi}=0,
\end{align}
where $\ket{\Psi}$ is any state created acting with $\sD^\dagger_\ad$ (also
with $\sA^\dagger_A$ and $\sA^\dagger_I$) on $\ket{E_-}$.

Later we will find an explicit description of these
states in terms of unintegrated vertex operators.

\section{Supergeometry}
\label{sec:supergeo}

The supergeometry is defined in terms of the coset element $g \in
\frac{CPSU(2,2|4)}{(\mathbb{R}^4\rtimes SO(4))\times
  (\mathbb{R}^4\rtimes SO(4))}$. We will parametrize $g$ with the
product of two factors. One depending only on $\sS_a$ and $\bar\sS_a$
and other with the remaining coset directions. The latter will be
\begin{align}
  g(x^+,x^-,x^A,x^I,\theta^\ad,\bar\theta^\ad)= e^{\im x^+\sE_+}
  e^{\im(x^-\sE_- +x^A\sP_A+x^I\sP_I+\theta^\ad\sS_\ad+\bar\theta^\ad \bar\sS_\ad)}.
\end{align}
Note that the generators in the second factor above almost form an
abelian algebra, the only non-trivial commutator is between the supercharges.
Then we have that
\begin{align}
  e^{-\im(x^-\sE_- +x^A\sP_A+x^I\sP_I+\theta^\ad\sS_\ad+\bar\theta^\ad \bar\sS_\ad)} d
  e^{\im(x^-\sE_- +x^A\sP_A+x^I\sP_I+\theta^\ad\sS_\ad+\bar\theta^\ad \bar\sS_\ad)} = \cr
  \im (dx^--\im d\theta^\ad \theta^\ad -\im d\bar\theta^\ad\bar\theta^\ad) \sE_- +
 \im  dx^A\sP_A +\im dx^I\sP_I + \im d\theta^\ad\sS_\ad +\im d \bar\theta^\ad \bar\sS_\ad.
\end{align}
The full current with $g(x^+,x^-,x^A,x^I,\theta^\ad,\bar\theta^\ad)$ will be
\begin{align}
  e^{-\im(x^A\sP_A+x^I\sP_I+\theta^\ad\sS_\ad+\bar\theta^\ad \bar\sS_\ad)} \im dx^+\sE_+
  e^{\im(+x^A\sP_A+x^I\sP_I+\theta^\ad\sS_\ad+\bar\theta^\ad \bar\sS_\ad)}\cr
  +\im (dx^--\im d\theta^\ad \theta^\ad -\im d\bar\theta^\ad\bar\theta^\ad) \sE_- +
 \im  dx^A\sP_A +\im dx^I\sP_I + \im d\theta^\ad\sS_\ad +\im d \bar\theta^\ad \bar\sS_\ad.
\end{align}
Because of the algebra, the expansion of the first term above stops in the terms quadratic in the fields
\begin{align}\label{smallGeo}
&J(x^+,x^-,x^A,x^I,\theta^\ad,\bar\theta^\ad)  =\cr
&~~~~~~~+\im \left(dx^- -\im d\theta^\ad
  \theta^\ad -\im d\bar\theta^\ad\bar\theta^\ad
  +dx^+\Big(\frac14 x^Ax^A +\frac14 x^Ix^I + \im\t_{\ad}\bar\t_{\bd}\Pi_{\ad\bd}\Big)\right) \sE_- \cr
&~~~~~~~+\im dx^+  \sE_+ +
  \im dx^A\sP_A +\im dx^I\sP_I +
  \im\big(d\theta^\ad +  dx^+\bar\theta^\bd\Pi_{\bd\ad}\big)\sS_\ad
+\im\big(d \bar\theta^\ad - dx^+ \theta^\bd\Pi_{\bd\ad}\big)
\bar\sS_\ad
\cr
&~~~~~~~ -\im dx^+ x^A \bar\sP_A -\im dx^+ x^I\bar\sP_I.
\end{align}
The first terms are the vielbein and last two terms are the connections.
This parametrization is the one that gives the usual metric in pp-wave
backgrounds. This is not completely true because we still have to
include $\theta^a$ and
$\bar\theta^a$. Since the algebra with the corresponding generators
only involves $\sE_-$, $\sS_a$, $\bar\sS_a$, $\sM_{AB}$ and $\sM_{IJ}$
we have that
\begin{align}
  &g^{-1}(\theta^a,\bar\theta^a) d g(\theta^a,\bar\theta^a) =\cr
  &(d\theta^a\theta^a +d\bar\theta^a\bar\theta^a)\sE_+ +\im E_0^a(\theta,\bar\theta)\sS_a
  +\im \bar E_0^a(\theta,\bar\theta)\bar\sS_a +
  \frac12\Omega_0^{AB}(\theta,\bar\theta)M_{AB}+\frac12\Omega_0^{IJ}(\theta,\bar\theta)\sM_{IJ}.
\end{align}
where the differential forms $(E^a,\bar E^a,\Omega^{AB},\Omega^{IJ})$ are
constrained by the Maurer-Cartan identity $dJ+J\wedge J=0$. Since
this sub-algebra is symmetric under the switch
$\sS_a \leftrightarrow \bar\sS_a$ we have that $\bar E^a_0(\theta,\bar\theta)=
E^a_0(\bar\theta,\theta)$ and $\Omega_0(\theta,\bar\theta)=
\Omega_0(\bar\theta,\theta)$. The Maurer-Cartan identities in components are
\begin{align}
  &d\theta^a\wedge d\theta^a+d\bar\theta^a\wedge d\bar\theta^a +
  E^a_0\wedge E^a_0+ \bar E^a_0\wedge \bar E^a_0 =0,\\
  &d E^a_0 + E^b_0\wedge \Omega_0^{AB}\sigma_{AB}^{ba}+
  E^b_0\wedge \Omega_0^{IJ}\sigma_{IJ}^{ba} =0,\\
  &d \bar E^a_0 + \bar E^b_0\wedge \Omega_0^{AB}\sigma_{AB}^{ba}+
  \bar E^b_0\wedge \Omega_0^{IJ}\sigma_{IJ}^{ba}=0,\\
  &d\Omega_0^{AB} +\Omega_0^{CD}\wedge\Omega_0^{EF}f^{AB}_{CD~EF}=0,\\
  &d\Omega_0^{IJ} +\Omega_0^{KL}\wedge\Omega_0^{MN}f^{IJ}_{KL~MN}=0,
\end{align}
where $f^{AB}_{CD~EF}$ and $f^{IJ}_{KL~MN} $ are $\mathfrak{so}(4)$ structure
constants. The explicit expression of these forms can be
found \cite{Metsaev:2001bj}, however we hopefully will not need them.

The have to full geometry we multiply the two coset factors as
\begin{align}
  g = g(\theta^a,\bar\theta^a) g(x^+,x^-,x^A,x^I,\theta^\ad,\bar\theta^\ad),
\end{align}
and the full Maurer-Cartan current is
\begin{align}
 g^{-1}dg=~& J(x^+,x^-,x^A,x^I,\theta^\ad,\bar\theta^\ad)+\cr &e^{-\im(x^A\sP_A+x^I\sP_I+\theta^\ad\sS_\ad+\bar\theta^\ad \bar\sS_\ad)}g^{-1}(\theta^a,\bar\theta^a) d g(\theta^a,\bar\theta^a)e^{\im(x^A\sP_A+x^I\sP_I+\theta^\ad\sS_\ad+\bar\theta^\ad \bar\sS_\ad)},
\end{align}
the expansion of the second term stops at quadratic order
in $(x^A,x^I,\theta^\ad,\bar\theta^\ad)$.

The expansion is
\begin{align}
g^{-1}dg=~  &\im J^+\sE_++\im J^-\sE_-+ \im J^A\sP_A+\im J^I\sP_I+\im
    J_1^a\sS_a+ \im J_1^\ad\sS_{\ad}\cr
    &+\im J_3^{a}\bar\sS_{a}+\im
    J_3^{\ad}\bar\sS_{\ad}+\frac12\O^{AB}\sM_{AB}+
    \frac12\O^{IJ}\sM_{IJ}+i\O^A\bar\sP_A+i\O^I\bar\sP_I ,
\label{Jcomp}
\end{align}
where
\begin{align}
&J^+=dx^+-\im(d\t_a\t_a+d\bar\t_a\bar\t_a) ,\\
&J^-=dx^--\im(d\theta_\ad
  \theta_\ad +d\bar\theta_\ad\bar\theta_\ad)\cr
&~~~~~~~  +\left(dx^+ -\im(d\t_a\t_a+d\bar\t_a\bar\t_a)\right) \left(\frac14 x^Ax^A +\frac14 x^Ix^I +\im \t_{\cd}\bar\t_{\dd}\Pi_{\cd\dd}\right)\cr
&~~~~~~~  +\frac{\im}{2}\left( (E_0X\Pi\bar\theta) - (\bar E_0X\Pi\theta)\right)-
  \frac{\im}{8}\left( (\theta\dot\Omega\theta) +
  (\bar\theta\dot\Omega\bar\theta)\right),\\
&J^A=dx^A-\im( E_0\sigma^A\theta +\bar E_0\sigma^A\bar\theta) -\O_0^{AB}X^B,\\
&J^I=dx^I-\im( E_0\sigma^I\theta +\bar E_0\sigma^I\bar\theta) -\O_0^{IJ}X^J,\\
&J_1^\ad=d\t^\ad-\frac14 (\dot\Omega\theta)^\ad + \left(dx^+ -
  \im(d\t^a\t^a+d\bar\t^a\bar\t^a) \right)   (\Pi\bar\theta)^\ad
  +\frac{1}{2}(\bar E_0X\Pi)^\ad  ,\\
&J_3^\ad= d\bar\t^\ad-\frac14 (\dot\Omega\bar\theta)^\ad -\left(  dx^+ - \im(d\t^a\t^a+d\bar\t^a\bar\t^a)   \right) (\Pi\theta)^\ad
  -\frac{1}{2}( E_0X\Pi)^\ad ,\\
&\O^A=-x^A\left(dx^+-\im(d\t^a\t^a+d\bar\t^a\bar\t^a)\right)-\im
  \left(E_0\sigma^A\Pi\bar\t-\bar E_0\sigma^A\Pi\t \right) ,\\
&\O^I=-x^I\left(dx^+-\im(d\t^a\t^a+d\bar\t^a\bar\t^a)\right)-\im
 \left(E_0\sigma^I\Pi\bar\t-\bar E_0\sigma^I\Pi\t \right),\\
&J_1^a= E_0^a,\quad J_3^a= \bar E_0^a ,\quad \O^{AB}=\O_0^{AB},\quad \O^{IJ}=\O_0^{IJ},
\end{align}
where we are using the following notations to have more compact
expressions
\begin{align}
  X_{a\ad}= x_A\sigma^A_{a\ad} + x_I \sigma^I_{a\ad}, \quad
  \dot\Omega_{\ad\bd}=
  \Omega_0^{AB}(\sigma_{AB})_{\ad\bd}+\Omega_0^{IJ}(\sigma_{IJ})_{\ad\bd},\\
  E_0^a(\sigma_A)_{a\ad}\theta^\ad = (E_0\sigma_A\theta),\quad
  E_0^a(\sigma_A\Pi)_{a\ad}\bar\theta^\ad =
  (E_0\sigma_A\Pi\bar\theta),\\
  \bar E_0^a(\sigma_A)_{a\ad}\bar\theta^\ad = (\bar E_0\sigma_A\bar\theta),\quad
  \bar E_0^a(\sigma_A\Pi)_{a\ad}\theta^\ad = (\bar E_0\sigma_A\Pi\theta),
\end{align}
and similar expressions.

The currents above define the whole supergeometry of the plane wave
background. They define the frame fields $dZ^M E_M{}^\bullet$, where
$\bullet$ is any of the $\frakg_1$, $\frakg_2$ or $\frakg_3$ directions and
connections $dZ^M\Omega_M{}^{\circ}$ where $\circ$ is any of the
$\frakg_0$ directions. The index $M$ is a local coordinate index for
the coset element $g(Z^M)$. In
our explicit parametrization this is identified with the $\bullet$ indices
. The covariant derivatives are defined with
the inverse of $E_M{}^\bullet$ and $\Omega_M{}^{\circ}$
\begin{align}
  \nabla_\bullet = E_\bullet{}^M\left( \partial_M -\Omega_M{}^\circ \sM_\circ\right).
\end{align}

For now we will restrict to the supergeometry of the coset
$(H(8|8)\rtimes U(1))/
(\mathbb{R}^4\times\mathbb{R}^4)$. From (\ref{smallGeo}) we can read off
$E_M{}^\bullet$ and $\Omega_M{}^\circ$ for this case and write the
covariant derivatives after inverting the vielbein
\begin{align}
  &\nabla_-=\partial_-,\quad \nabla_+=\partial_+ -\frac{1}{4}x^2 \partial_-
  -\bar\theta_\ad\Pi_{\ad\bd}\partial_\bd
  +\theta_\ad\Pi_{\ad\bd}\bar\partial_\bd +\im x^A\bar\sP_A+\im x^I\bar\sP_I , \\
  &\nabla_A=\partial_A,\quad \nabla_I=\partial_I,\quad \nabla_\ad
  =\partial_\ad +\frac{\im}{2}\theta_\ad\partial_-,\quad
  \bar\nabla_\ad = \bar\partial_\ad +\frac{\im}{2}\bar\theta_\ad\partial_-,\\
  &\bar\nabla_A = \bar\sP_A,\quad
  \bar\nabla_I = \bar\sP_I,
\end{align}
where $(\bar\nabla_A,\bar\nabla_I)$ are the generators of the boosts
in the directions $A$ and $I$. The $(\bar\sP_A,\bar\sP_I)$ should be
understood as acting on coset elements $g$ by multiplication from the
right. Therefore they are defined to satisfy
\begin{align}
  &[\nabla_+,\bar\sP_A] = \im \nabla_A,\quad [\nabla_+,\bar\sP_I] =
  -\im \nabla_I,\\
  &[\nabla_A,\bar\sP_B] =\frac{\im}{2}\delta_{AB}\nabla_-,\quad
  [\nabla_I,\bar\sP_J] =-\frac{\im}{2}\delta_{IJ}\nabla_-.
\end{align}
Furthermore, the isotropy generators do not act on the
coordinates. This is because there is no linear variation of
coordinates such that $g^{-1}\delta g=v^A \bar\sP_A +v^I\bar\sP_I$.
It can be verified that the algebra of these covariant
derivatives is the same as their corresponding generators but with an
extra $-\im$ multiplying the structure constants. This is because, by
definition, the covariant derivatives are differential operators
such that when acting in the coset element $g$ we have that
\begin{align}
  \nabla_\bullet g = \im g\sT_\bullet,
\end{align}
where $\sT_\bullet$ is the corresponding algebra generator. Similarly,
the isometry generators can also be represented as differential
operators with the property that
\begin{align}\label{symmetryGen}
  {\mathsf t}_\bullet g = \sT_\bullet g.
\end{align}
The reason for the different hermiticity conventions is that usually
one wants covariant derivatives that are anti-hermitian, but symmetry
generators are usually hermitian.
The expressions for the symmetry generators are
\begin{align}
  &\se_- = -\im\partial_-=,\quad \se_+ =-\im\partial_+, \\
  &\sp_A=- \im\cos(x^+)\partial_A +\sin(x^+)\bar\sP_A
  +\frac{\im}{2}\sin(x^+)x_A\partial_-,\\
  &\bar\sp_A = \cos(x^+)\bar\sP_A +\im\sin(x^+)\partial_A
    +\frac{\im}{2}\cos(x^+)x_A\partial_-\\
   &\sp_I= -\im\cos(x^+)\partial_I +\sin(x^+)\bar\sP_I
  +\frac{\im}{2}\sin(x^+)x_I\partial_-,\\
  &\bar\sp_I =  \cos(x^+)\bar\sP_I +\im\sin(x^+)\partial_I
    +\frac{\im}{2}\cos(x^+)x_I\partial_-\\
  &\ss_\ad = -\im\cos\left(x^+\right)\sq_\ad  -\im
    \sin\left(x^+\right)\Pi_{\ad\bd}\bar\sq_\bd,\\
  &\bar\ss_\ad = -\im\cos\left(x^+\right)\bar\sq_\ad +
    \im\sin\left(x^+\right)\Pi_{\ad\bd}\sq_\bd,
\end{align}
where
\begin{align}
  \sq_\ad = \partial_\ad
  -\frac{\im}{2}\theta_\ad\partial_-, \quad
  \bar\sq_\ad = \bar\partial_\ad
    -\frac{\im}{2}\bar\theta_\ad\partial_-.
\end{align}

The operators $(\bar\sP_A,\bar\sP_I)$ above are the same as the ones
used in the covariant derivative, so it should be understood
as acting on $g$ from the right. This also means they commute
with all the partial derivatives above. Using this we
can check that
\begin{align}
  [\se_+,\sp_A]= -\im\bar\sp_A,\quad [\sp_A,\bar\sp_B]=\frac{\im}{2}\se_-,\quad [\se_+,\bar\sp_A]=\im \sp_A,\quad {\rm etc.}
\end{align}
The origin of the additional minus sign is from the definition (\ref{symmetryGen})
\begin{align}
  \st_1\st_2 g= \st_1 \sT_2g=\sT_2\st_1g=\sT_2\sT_1g.
\end{align}

If we include dependence on the remaining odd directions there
will be further contributions to all operators above. In particular we
note that $(\bar\sp_A,\bar\sp_I)$ will get contributions like
\begin{align}\label{RchargeRaise}
  &\bar\sp_A(\theta^a,\bar\theta^a)= \bar\sp_A^{(0)} -\frac{1}{2} (\theta\sigma_A)_\ad
 \ss_\ad^{(0)} -\frac{1}{2}(\bar\theta\sigma_A)_\ad \bar\ss_\ad^{(0)} +\cdots,\\
  &\bar\sp_I(\theta^a,\bar\theta^a) =\bar\sp_I^{(0)} -\frac{1}{2} (\theta\sigma_I)_\ad \ss_\ad^{(0)}
  -\frac{1}{2}(\bar\theta\sigma_I)_\ad\bar\ss_\ad^{(0)} +\cdots
\end{align}
\begin{align}\label{RchargeRaise2}
  &\sp_A(\theta^a,\bar\theta^a)= \sp_A^{(0)} +\frac{1}{2} (\theta\sigma_A\Pi)_\ad
 \bar\ss_\ad^{(0)} -\frac{1}{2}(\bar\theta\sigma_A\Pi)_\ad \ss_\ad^{(0)} +\cdots,\\
  &\sp_I(\theta^a,\bar\theta^a) =\sp_I^{(0)} +\frac{1}{2} (\theta\sigma_I\Pi)_\ad \bar\ss_\ad^{(0)}
  -\frac{1}{2}(\bar\theta\sigma_I\Pi)_\ad\ss_\ad^{(0)} +\cdots
\end{align}
Similarly, the operators $(\ss_\ad,\bar\ss_\ad)$ get the contributions
\begin{align}\label{RSS}
\ss_\ad(\t^a,\bar\t^a)=&~~  \ss_\ad^{(0)} - \im (\t\s_i)_\ad \sp_i^{(0)}
 + \im (\bar\t\s_i\Pi)_\ad\bar\sp_i^{(0)}  +\cdots \\ \label{RSSbar}
  \bar\ss_\ad(\t^a,\bar\t^a) =&~~  \bar \ss_\ad^{(0)} -
  \im (\bar\t\s_i)_\ad \sp_i^{(0)} - \im (\t\s_i\Pi)_\ad\bar\sp^{(0)}_i +\cdots
\end{align}
We will use these expressions later. It will also be important that
the expression for the differential generator $\se_+$ does not change
in the full coset
\begin{align}
  \se_+ = \se^{(0)}_+ = -\im \partial_+. \label{fulleplus}
\end{align}

\section{Action, BRST and conformal invariance}
\label{sugra}

In  this section we will review the sigma model action for the pure
spinor string in the plane wave background \cite{Berkovits:2002zv}\
and prove its invariance under the BRST-like
transformations. As in the $AdS_5\times S^5$ case, the geometric part
of the action is constructed using the Maurer-Cartan one-form
$J=g^{-1}dg$, where $g$ is a coset element. This one-form is expanded
in the algebra elements as
\begin{align}
J=~  &\im J^+\sE_++\im J^-\sE_-+ \im J^A\sP_A+\im J^I\sP_I+\im
    J_1^a\sS_a+ \im J_1^\ad\sS_{\ad}\cr
    &+\im J_3^{a}\bar\sS_{a}+\im
    J_3^{\ad}\bar\sS_{\ad}+\frac12\O^{AB}\sM_{AB}+
    \frac12\O^{IJ}\sM_{IJ}+i\O^A\bar\sP_A+i\O^I\bar\sP_I .
\label{Jcomp}
\end{align}
The world-sheet action is
\begin{align}
S=& \int d^2z\Big( - J^+\bar J^- -  J^-\bar J^+ +\frac12
    J^A \bar J^A +\frac12 J^I \bar J^I +d_a \bar J_1^a + d_{\ad}\bar
    J_1^{\ad} +\bar d_a J_3^{a} +\bar d_{\ad} J_3^{\ad} \cr
    & +\frac{\im}{2} \Pi^{\ad\bd}d_{\ad}\bar d_{\bd}
    +\o_a\bar\nabla\l^a+\o_{\ad}\bar\nabla\l^{\ad}
    +\bar\o_{a}\nabla\bar\l^{a}+\bar\o_{\ad}\nabla\bar\l^{\ad}-
     N_A \bar N_A-N_I\bar N_I \Big) + S_{\rm WZ}.
\label{action}
\end{align}
This action uses the following definitions. The fields
$(\l^a,\l^\ad,\bar\l^a,\bar\l^\ad)$  are the pure spinor ghosts that
satisfy
\begin{align}
  \l^a\sigma^A_{a\ad}\l^\ad=\l^a\sigma^I_{a\ad}\l^\ad=\bar\l^a\sigma^A_{a\ad}\bar\l^\ad=
  \bar\l^a\sigma^I_{a\ad}\bar\l^\ad=\l^a\l^a=\l^\ad\l^\ad=\bar\l^a\bar\l^a
  =\bar\l^\ad\bar\l^\ad=0
\end{align}
The fields $(\o_a,\o_\ad,\bar\o_a,\bar\o_\ad )$ are their conjugate
momenta. The covariant derivatives are
\begin{align}\label{covDevGhosts}
  &\bar\nabla \l^a = \bar\partial \l^a - \frac12 \bar\Omega^{ab}\l^b,\quad
  \nabla\bar\l^a =  \partial \bar\l^a - \frac12 \Omega^{ab}\bar\l^b,\\
  &\bar\nabla \l^\ad = \bar\partial \l^\ad - \frac12
  \bar\Omega^{\ad\bd}\l^\bd -\frac12\bar\Omega^{\ad a}\l^a,\quad
  \nabla\bar\l^\ad =  \partial \bar\l^\ad - \frac12 \Omega^{\ad\bd}\bar\l^\bd
  -\frac12\Omega^{\ad a}\bar\l^a,
\end{align}
where
\begin{align}
&\Omega^{ab}=\frac12\Omega^{AB}(\sigma_{AB})^{ab}+\frac12\Omega^{IJ}(\sigma_{IJ})^{ab},\quad
  \Omega^{\ad\bd}=\frac12\Omega^{AB}(\sigma_{AB})^{\ad\bd}+\frac12\Omega^{IJ}(\sigma_{IJ})^{\ad\bd},\\
  &\Omega^{\ad a} = \Omega^A (\sigma_A)^{a\ad}+\Omega^I (\sigma_I)^{a\ad},
\end{align}
and analogous expressions for the left-moving connections. The reason
for the asymmetry in the definitions of covariant derivatives is that
after the contraction, only the spinor with dot type index transforms
under boosts in the $A$ and $I$ directions. The new currents in the
second line are
\begin{align}
N_A=\frac12\l_a\o_{\bd}(\s_A)_{a\bd},\quad
  N_I=\frac12\l_a\o_{\bd}(\s_I)_{a\bd},\quad
  \bar N_A=\frac12\lb_{a}\ob_{\ad}(\s_A)_{a\ad},\quad
  \bar N_I=\frac12\lb_{a}\ob_{\ad}(\s_I)_{a\ad}.
\end{align}
We can also define
\begin{align}
  N^{AB}=\frac12 \l_a \o_b (\s^{AB})_{ab} +\frac12 \l_{\ad}
  \o_{\bd}(\s^{AB})_{\ad\bd},\quad
  N^{IJ}=\frac12 \l_a \o_b (\s^{IJ})_{ab} + \frac12 \l_{\ad} \o_{\bd}
  (\s^{IJ})_{\ad\bd},\\
   \bar N^{AB}=\frac12 \bar\l_a \bar\o_b (\s^{AB})_{ab} +
  \frac12 \bar\l_{\ad}\bar\o_{\bd}(\s^{AB})_{\ad\bd},\quad
  \bar N^{IJ}=\frac12 \bar\l_a \bar\o_b (\s^{IJ})_{ab} +
  \frac12 \bar\l_{\ad} \bar\o_{\bd} (\s^{IJ})_{\ad\bd}.
\end{align}
Note that using all these definitions, the kinetic terms for the ghosts
can be written as
\begin{align}
  \o_a\bar\nabla\l^a+\o_{\ad}\bar\nabla\l^{\ad}
  +\bar\o_{a}\nabla\bar\l^{a}+\bar\o_{\ad}\nabla\bar\l^{\ad} =
  \o_a\bar\partial\l^a+\o_{\ad}\bar\partial\l^{\ad}
  +\bar\o_{a}\partial\bar\l^{a}+\bar\o_{\ad}\partial\bar\l^{\ad}\cr
 - \frac12\bar\Omega^{AB}N_{AB}-
\frac12\Omega^{AB}\bar N_{AB}-\frac12\O^{IJ}\bar N_{IJ}-\frac12\bar\Omega^{IJ}N_{IJ}-\bar\Omega^A N_A
  -\bar\Omega^I N_I - \Omega^A\bar N_A -\Omega^I\bar N_I.
\end{align}

Finally, $S_{\rm WZ}$ is the Wess-Zumino term which is defined on a
three-dimensional surface whose boundary is  world-sheet of the
string  \cite{Berkovits:1999zq} and its most compact form is
\begin{align}
  S_{\rm WZ}=k \int_{\Sigma_3}  \gamma_{m\alpha\beta} \left(
 J^m_2\wedge J^\alpha_1\wedge J^\beta_1-
 J^m_2\wedge J^\alpha_3\wedge J^\beta_3\right) ,
\label{SWZ}
\end{align}
where $\gamma_{m\alpha\beta}$ is the $\gamma$-matrix in ten
dimensional notation. Unlike the $AdS_5\times S^5$ case, the WZ term
cannot be written as an integral of a globally defined two-form for
the plane wave background. Despite this fact, nevertheless, as usual
expected for a Wess-Zumino term,  any variation of $S_{WZ}$ can be
written as an integral at the boundary. If a variation along the
coset directions is  parametrized by $\e=g^{-1}\d g$, (\ref{SWZ}) transforms to
\begin{align}
  \int\!\! \left( \e^m_2 ( J^\alpha_1\wedge J^\beta_1 -
  J^\alpha_3\wedge J^\beta_3 )  +
  \e^\alpha_1 ( J^\beta_1\wedge J^m_2 + J^m_2\wedge J^\alpha_1 ) -
  \e^\alpha_3 ( J^\beta_3 \wedge J^m_2 + J^m_2\wedge J^\beta_3 )\!
  \right)\! \gamma_{m\alpha\beta}.
\label{WZvariation}
\end{align}

For the geometric part of the action, the BRST-like transformation
for the coset element  $g$ is $Qg=g(\l+\lb)$ where
$\l=\l^a\sS_a+\l^{\ad}\sS_{\ad}$
and $\lb=\lb^{a}\bar\sS_{a}+\lb^{\ad}\bar\sS_{\ad}$. The
transformation of $J=g^{-1}dg$ is $QJ=d(\l+\lb)+[J,\l+\lb]$.
Using the algebra from Section \ref{review} we obtain
\begin{align}
  &QJ^+=\l^a J_1^a+\lb^{a}J_3^{a},\quad
  QJ^-=\l^{\ad}J_1^{\ad}+\lb^{\ad}J_3^{\ad},\label{QJ2lc}\\
  & Q J_2^A=(\l\s^AJ_1)+(J_1\s^A\l)+(\bar\l\s^AJ_3)+(J_3\s^A\bar\l),\cr
  &Q J_2^I= (\l\s^IJ_1)+(J_1\s^I\l)+(\bar\l\s^IJ_3)+(J_3\s^I\bar\l)
  \label{QJ2transverse}\\
  & QJ_1^a=-\im\nabla\l^a,\quad
  QJ_1^{\ad}=-\im\nabla\l^{\ad}-\im J_2^+(\bar\l\Pi)^\ad+
  \frac{\im}{2}J_2^A(\bar\l\s_A\Pi)^\ad+ \frac{\im}{2}J_2^I(\bar\l\s_I\Pi)^\ad ,
  \label{QJ1}\\
  &QJ_3^{a}=-\im\nabla\lb^a, \quad
  QJ_3^{\ad}=-\im\nabla\lb^{\ad}+\im J_2^+(\l\Pi)^\ad-
    \frac{\im}{2} J_2^A(\l\s_A\Pi)^\ad- \frac{\im}{2} J_2^I(\l\s_I\Pi)^\ad .
\label{QJ3}
\end{align}
From the same calculation we also obtain the BRST-like transformations
of the connections
\begin{align}
&Q\O^{AB}=(\l^a J_3^b + \lb^b J_1^a) (\s^{AB}\Pi)_{ab} \cr
&Q\O^{IJ}=-(\l^a J_3^{b} + \lb^{b} J_1^a) (\s^{IJ}\Pi)_{ab} \cr
&Q\O^A=(\l^aJ_3^{\bd}+\lb^{\bd}J_1^a)(\s^A\Pi)_{a\bd}+
    (\l^{\ad}J_3^{b}+\lb^{b}J_1^{\ad})(\s^A\Pi)_{\ad b} \cr
&Q\O^I=(\l^aJ_3^{\bd}+\lb^{\bd}J_1^a)(\s^I\Pi)_{a\bd}-
    (\l^{\ad}J_3^{b}+\lb^{b}J_1^{\ad})(\s^I\Pi)_{\ad b}.
\end{align}
It remains to define the transformations of the fields not defined by
the geometry.  The pure spinor ghost variables $\l, \lb$ are BRST
invariant and the pure spinor antighosts transform as
\begin{align}
  Q \o_a=-\im d_a,\quad Q \o_{\ad}=-\im d_{\ad},\quad
  Q \ob_a=-\im \bar d_a, \quad Q \ob_{\ad}=-\im \bar d_{\ad} .
\label{Qom}
\end{align}
The last fields are the supersymmetric momenta
\begin{align}\label{Qda}
  Q d_a=&~ 2\l_a J_2^{-}- J^A (\s_A\l)_a-J^I (\s_I\l)_a +
        \frac12 (N^{AB} (\s_{AB}\Pi\lb)_a- N^{IJ}(\s_{IJ}\Pi\lb)_a) \cr
  &+(N_A(\s_A\Pi\lb)_{a}+N_I(\s_I\Pi\lb)_{a}) ,\\
 \label{Qdadot}
  Q d_{\ad}=&~ 2\l^{\ad}J^+-J^A_2(\l\s_A)_{\ad} -J_2^I(\l\s_I)_{\ad}+
        (N_A(\s_A\Pi\lb)_{\ad}-N_I(\s_I\Pi\lb)_{\ad} ) ,\\
  \label{Qdbara}
  Q \bar d_{a}= &~2\lb_a\bar J_2^{-} -\bar J_2^A (\s_A\lb)_{a}-\bar J_2^I (\s_I\lb)_{a}
       + \frac12(\bar N^{AB} (\l\s_{AB}\Pi)_a-\bar N^{IJ}(\l\s_{IJ}\Pi)_{a}) \cr
        &+(\bar N_A(\l\s_A\Pi)_{a}-\bar N_I(\l\s_I\Pi)_{a}),\\
\label{Qdbaradot}
  Q \bar d_{\ad}=&~ 2\lb_{\ad}\bar J_2^+-\bar J_2^A(\lb\s_A)_{\ad}-\bar J_2^I(\lb\s_I)_{\ad}
                 +(\bar N_A(\l\s_A\Pi)_{\ad}-\bar N_I(\l\s_I\Pi)_{\ad} ) ,
\end{align}

The calculation of the BRST transformation of the action can be
organized as follows. We first note that for the case of a BRST
transformation the $\epsilon$ in (\ref{WZvariation}) is given by
$\l^a\sS_a+\l^{\ad}\sS_{\ad}
+\lb^{a}\bar\sS_{a}+\lb^{\ad}\bar\sS_{\ad}$. In this case, it
simplifies to
\begin{align}
Q S_{WZ}=k\int d^2z\Big( \l^a(J^{[-}\bar J^{a]}-\frac12 J^{[i}\bar J^{\bd]}\s^i_{a\bd})+\l^{\ad}(J^{[+}\bar J^{\ad]}-\frac12 J^{[i}\bar J^{b]}\s^i_{b\ad}) \cr
-\lb^{a}(J^{[-}\bar J^{a]}-\frac12 J^{[i}\bar
  J^{\bd]}\s^i_{\a\bd} ) - \lb^{\ad}( J^{[+}\bar J^{\ad]}-\frac12
  J^{[i}\bar J^{b]}\s^i_{b\ad}) \Big).
\label{QSWZ}
\end{align}
The strange anti-symmetrization in different type of indices should
actually be read off as an anti-symmetrization of the left- and
right-moving currents. It turns out that if the constant  $k$ is equal
to $1$ the BRST transformation of $\int d^2z (-J^{(+}\bar
J^{-)}+\frac12 J^i\bar J^i) + S_{WZ}$ is
\begin{align}
\int d^2z \Big(& (-2\l^a J_2^-+\l^{\bd}J_2^A\s^A_{a\bd}+\l^{\bd}J^I\s^I_{a\bd})\bar
  J_1^a+(-2\l^{\ad}J^++\l^bJ_2^A\s^A_{b\ad}+\l^bJ_2^I\s^I_{b\ad})\bar
  J_1^{\ad}\\
 & +(-2\lb^{a}\bar J^-+\lb^{\bd}\bar
  J_2^A\s^A_{a\bd}+\lb^{\bd}\bar
   J_2^I\s^I_{a\bd})J_3^{a}+
   (-2\lb^{\ad}\bar J_2^++\lb^{b}\bar
  J_2^A\s^A_{b\ad}+\lb^{b}\bar
  J_2^I\s^I_{b\ad})J_3^{\ad} \Big) .
\end{align}

This expression is canceled by some terms in BRST transformations of fields
$(d_a,d_\ad,\bar d_a,\bar d_\ad)$ in  $\int( d_a \bar J_1^a + d_{\ad}\bar
J_1^{\ad} +\bar d_a J_3^{a} +\bar d_{\ad} J_3^{\ad}+
\frac{\im}{2} \Pi^{\ad\bd}d_{\ad}\bar d_{\bd})$. The terms from $Qd$ that remain,
together with the transformations of $(\bar J_1^a,\bar
J_1^\ad,J_3^a,J_3^\ad)$ will cancel with the transformations of the
anti-ghosts and connections in the ghost part of the action.

It was argued in \cite{Berkovits:2002zv}\ that the action
(\ref{action}) is conformally invariant to all orders in perturbation
theory. The argument goes as follows. Using the supergeometry defined
in Section \ref{sec:supergeo} we can calculate the explicit form of
the action in terms of the parametrization for $g$. If we assign a
positive charge to $(\theta^a,\bar\theta^a)$ and a negative charge to
$(d_a,\bar d_a)$ we can separate the action in two parts. One part has positive
charge and the other has zero charge. We will call this $S$-charge.
This comes from the expansion of the Maurer-Cartan currents
$(J^\bullet,J^\circ)$.
The part that has zero $S$-charge contains
the kinetic term for $(\theta^a,d_a,\bar\theta^a,\bar d_a)$, the
ghosts and a coset sigma model generated by
$\{\sE_+,\sE_-,\sP_A,\sP_I,\sS_\ad,\bar\sS_\ad,\bar\sP_A,\bar\sP_I\}/
\{\bar\sP_A,\bar\sP_I\}$. Because of the structure of the vertices in
the coset sigma model and ghosts, the divergent part of the
effective action of the zero $S$-charge part has only a one
loop contribution, which vanishes.  Since
the zero $R$-charge part is tied to the positive charge part by the isometry
transformations and that the propagator
for $(\theta^a,d_a,\bar\theta^a,\bar d_a)$ conserves the $S$-charge, it
follows that the whole action is conformally invariant to all loop
orders. Only the full Maurer-Cartan current $(J^\bullet,J^\circ)$ is
invariant under all isometries. This drastic simplification comes
from the fact there is no
Ramond-Ramond flux coupling the fields $(d_a,\bar d_a)$ with the rest
of the variables. In the following section we will argue that, at
least for the massless sector,  the
physical spectrum can be found by looking at unintegrated vertex
operators with zero $S$-charge.

\section{Massless vertex operators}
\label{sec:Masslessvertex}

Vertex operators in string theory comes in two flavors, unintegrated
and integrated. They describe the same spectrum and both are need to
compute observables. In the pure spinor formalism they are related by a
chain of equations \cite{Grassi:2004ih} that follows from the
BRST-like symmetry of the theory. The integrated vertex are
interpreted as deformations of the action.

The unintegrated vertex operators are space-time scalars with
conformal vanishing world-sheet conformal dimension. For massless
states this means they are constructed with world-sheet scalars and
its anomalous dimension has to vanish. Their general form is
$U\big(\l^a,\bar\l^a,\l^\ad,\bar\l^\ad, g(Z^M)\big)$. The cohomology
defined by the BRST-like transformations implies that at ghost number
zero the only physical operator is the identity. At ghost number one,
the cohomology is found to be related to the conserved currents
corresponding to the space-time global symmetries \cite{Mikhailov:2009rx}.
The massless spectrum is in the ghost number two cohomology. Using the
$\mathfrak{so}(4)\oplus\mathfrak{so}(4)$ notation we write
\begin{align}
  U\big(\l^a,\bar\l^a,\l^\ad,\bar\l^\ad, g(Z^M)\big)=&~
  \l^a\bar\l^bU_{ab}\big(g(Z^M)\big) + \l^a\bar\l^\bd
  U_{a\bd}\big(g(Z^M)\big)\cr& + \l^\ad\bar\l^b U_{\ad b}\big(g(Z^M)\big) +
  \l^\ad \bar\l^\bd U_{\ad\bd}\big(g(Z^M)\big).
\end{align}

The physical state conditions comes from the condition that
$U\big(\l^a,\bar\l^a,\l^\ad,\bar\l^\ad, g(Z^M)\big)$ is invariant
under the BRST-like transformations. Since the ghosts are invariant,
the only contribution comes from the coset element $\delta g=
g(\l^a\sS_a+\l^\ad\sS_\ad+\bar\l^a\bar\sS_a+\bar\l^\ad\bar\sS_\ad)$.
For a general function of $g$, we have that
\begin{align}
  f(g+\delta g) = f(g) +
  (\l^a\nabla_a+\l^\ad\nabla_\ad+\bar\l^a\bar\nabla_a+\bar\l^\ad\bar\nabla_\ad)f(g),
\end{align}
where $(\nabla_a,\nabla_\ad,\bar\nabla_a,\bar\nabla_\ad)$ are the
covariant derivatives defined in Section \ref{sec:supergeo}.

As in the case of the action, we can expand a general function $f(g)$ of the
coset in $S$-charge powers
\begin{align}
  f(g) =\sum_{n=0}^{16} f^{(n)}(g).
\end{align}
The term with zero $S$-charge is a function of only the smaller
coset $(H(8|8)\rtimes U(1))/(\mathbb{R}^4\times\mathbb{R}^4)$.
We will denote an element of the this coset
by $g_0$. So we have that
\begin{align}
   f(g) = f(g_0)+ \sum_{n=1}^{16} f^{(n)}(g),
\end{align}

If the function is a space-time scalar it must be invariant under all
isometries, in particular it must be invariant under the isometries
generated by $(\bar\sp_A,\bar\sp_I)$
\begin{align}
  f(g')= f(\Lambda g)= f(g),
\end{align}
where $\Lambda$ is a finite isometry transformation. From the algebra
we can see that, for example, the isometry transformations along the directions
$\{\bar\sp_A,\bar\sp_I\}$ raise the $S$-charge. It follows from
(\ref{RchargeRaise}) that
\begin{align}
  \delta_A\theta^a =0,\quad \delta_A\theta^\ad =
  \frac{\im}{2}(\theta\sigma_A)^\ad.
\end{align}
The consequence of this is that all different $S$-charge
powers of $f(g)$
are tied together by global isometry invariance.
This is very similar to the argument used
to prove conformal invariance of the full action once the vanishing
$S$-charge part was found to be finite
\cite{Berkovits:2002zv}. Therefore we will first find what is
appropriate vertex operator starting with functions of the smaller
coset $f(g_0)$.

Imposing that $U$ must be invariant under all isometries is too
strong. For example, in flat space, if we demand that a vertex
operator is invariant under all translations its momentum should
vanish. For the coset, the only possible invariant combinations that
are invariant under all isometries are the differentials $g^{-1}dg$
and finite differences $g^{-1}_1g_2$.
Later in this section we will discuss what are the
appropriate conditions to impose on $U$ such that we find its full
superspace form.

Suppose we can find a single  vertex operator
$U(x^-,x^+,x^A,x^I,\theta^\ad,\bar\theta^\ad,\lambda,\bar\lambda)$
corresponding to a scalar that
satisfy $Q U=0$ and that depends only on a scalar polarization that are
invariant under some of the isometries. If such operator exists
we can construct the full $U$
order by order in $(\theta^a,\bar\theta^a)$. It is crucial that the
generators $(\bar\sP_A,\bar\sP_I)$ inside the differential form of the
isometry generators act only on the ghosts and polarizations.

We will now construct a scalar vertex operator. First we start with an
ansatz that depends only on $(x^-,\theta^\ad,\bar\theta^\ad)$.
The general form of the vertex operator will be
\begin{align}
  U_0(x^-,\theta^\ad,\bar\theta^\ad,\lambda,\bar\lambda) =&
  \lambda^a\bar\lambda^b U_{ab}(x^-,\theta^\ad,\bar\theta^\ad) +
  \lambda^\ad\bar\lambda^b U_{\ad b}(x^-,\theta^\ad,\bar\theta^\ad)+\\
  &\lambda^a\bar\lambda^\bd U_{a\bd}(x^-,\theta^\ad,\bar\theta^\ad)+
  \lambda^\ad\bar\lambda^\bd U_{\ad\bd}(x^-,\theta^\ad,\bar\theta^\ad).
\end{align}
Two of the equations that come from BRST invariance are
\begin{align}
  \nabla_\ad U_{\bd a} + \nabla_{\bd}U_{\ad a} =\delta_{\ad\bd}A_a,\quad
  \bar\nabla_\ad U_{a \bd} + \bar\nabla_\bd U_{a \ad} =
  \delta_{\ad\bd} \bar A_a.
\end{align}
If $\nabla_-
U(x^-,\theta^\ad,\bar\theta^\ad,\lambda,\bar\lambda)\neq 0$, the case
where this vanishes will be seen later, these
equations can be solved as
\begin{align}
  U_{\ad a} = \frac{\nabla_\ad}{2\nabla_-} A_a,\quad
  U_{a\ad}=\frac{\bar\nabla_\ad}{2\nabla_-} \bar A_a,
\end{align}
but these can be canceled by a gauge transformation for the
vertex. Since the component $U_{\ad\bd}$ is related with the ones
above by the $(\bar\sP_A,\bar\sP_I)$ isotropies, it will also
vanish. Only $(\lambda^a,\bar\lambda^a)$ are invariant under these isotropies.
The remaining equations are
\begin{align}
  \nabla_\ad U_{a b} =(\sigma^A)_{a\ad} U^A_b+(\sigma^I)_{a\ad}U^I_b,
  \quad \bar\nabla_\ad U_{ab} =(\sigma^A)_{b\ad}\bar U^A_a +
  (\sigma^I)_{b\ad}\bar U^I_a.
\end{align}
These equations can be solved with two known superspace functions
$f_a(E_-,\theta^\ad,\eta^A,\eta^I,\eta_a)$ and
$\bar f_a(E_-,\bar\theta^\ad,\bar\eta^A,\bar\eta^I,\bar\eta_a)$ that
satisfy
\begin{align}
  \left(\partial_\ad -\frac{E_-}{2}\theta_\ad\right) f_a =
  (\sigma^A)_{a\ad}f^A +(\sigma^I)_{a\ad} f^I,\quad
  \left(\bar\partial_\ad -\frac{E_-}{2}\bar\theta_\ad\right)\bar f_a = (\sigma^A)_{a\ad}f^A +(\sigma^I)_{a\ad} \bar f^I,
\end{align}
that depend on the set of polarizations
$(\eta^A,\eta^I,\eta_a,\bar\eta^A,\bar\eta^I,\bar\eta_a)$. Their
explicit expressions can be found in \cite{Brink:1983pf,
  Jusinskas:2014vqa, Berkovits:2014bra}. Up to
now our ansatz for the  unintegrated vertex operator is
\begin{align}
  U_0(x^-,\theta^\ad,\bar\theta^\ad,\lambda^a,\bar\lambda^a) =
  \lambda^a\bar\lambda^b f_a(E_-,\theta^\ad,\eta^A,\eta^I,\eta_a)
\bar f_a(E_-,\bar\theta^\ad,\bar\eta^A,\bar\eta^I,\bar\eta_a) e^{\im
  E_- x^-}.
\end{align}
It should be noted that $x^-$ is not a periodic variable so $E_-$ can
have any real value. The functions $(f_a,\bar f_a)$ are singular
in the limit $E_-\to 0$. However, in the plane wave background
we can construct a scalar superfield
that is well defined in this limit
\begin{align}
  \Phi(E_-,\eta^A,\eta^I,\eta^a,\bar\eta^A,\bar\eta^I,\bar\eta^a)=
   (E_-)^2 \Pi^{ab}f_a(E_-,\theta^\ad,\eta^A,\eta^I,\eta_a)
   \bar f_a(E_-,\bar\theta^\ad,\bar\eta^A,\bar\eta^I,\bar\eta_a).
\end{align}
In the limit $E_-\to 0$ the superfield reduces to
\begin{align}
  \Phi(0,\eta^A,\eta^I,\eta^a,\bar\eta^A,\bar\eta^I,\bar\eta^a)= 4\Pi^{ab}\eta_a\bar\eta_b,
\end{align}
The scalar $\Phi$ is the field  that changes the background value of the Ramond-Ramond flux by a constant amount \cite{Berkovits:2008ga,Chandia:2017afc}. However for $E_-\neq 0$ it is not a scalar since it depends on the polarizations inside
the functions $(f_a,\bar f_a)$. In order to have a scalar we will set
\begin{align}
  &\eta^A=\eta^I=\bar\eta^A=\bar\eta^I=0,\\
  & \eta_a\bar\eta_b =  (E_-)^2\Pi_{ab}.
\end{align}

This initial ansatz for the unintegrated vertex operator is then
\begin{align}
&U_0(x^-,\theta^\ad,\bar\theta^\ad,\lambda^a,\bar\lambda^a)=\cr
&~~~~~~~~~~~~~~~~~~~\lambda^a\bar\lambda^a f_a(E_-,\theta^\ad,0,0,\eta_a)
    \bar f_a(E_-,\bar\theta^\ad,0,0,\bar\eta_a)e^{\im E_- x^-}
    \Big|_{\eta_a\bar\eta_b=(E_-)^2\Pi_{ab}}.
\end{align}
In the expression above we have set $\phi=1$. We will include it
explicitly when discussing the construction of the massless spectrum.

The next step is to introduce dependence on the remaining
bosonic coordinates. Since we are working with the smaller coset we
are missing the constraints imposed by
$\lambda^a\nabla_a+\bar\lambda^a\bar\nabla_a$. In particular, the full
Virasoro constraint cannot be obtained if this part of the BRST
transformation is not included. We will take another route and impose
that the vertex operator is killed by the Casimir operator
$\mathfrak{C}_2$ (\ref{casimirH88}) in differential form. The
consequence is that one effectively imposes $L_0 +\bar L_0 =0$ on
the state. This can be done using the covariant derivatives or the
symmetry generators. The quadratic Casimir evaluated with both set of
operators differ by a minus sign due to the different hermiticity
convention
\begin{align}
  \mathfrak{C}_2 = -\partial_+\partial_- +\frac{x^2}{4}\partial^2_- +
   \partial_A\partial_A +\partial_I\partial_I -\im\Pi^{\ad\bd}\ss_\ad\bar\ss_\bd.
\end{align}
Since the vertex operators are invariant under isotropy
transformations we drop $(\bar\sP_A,\bar\sP_I)$. Notice that the
quadratic Casimir can be written as
\begin{align}
  \mathfrak{C}_2 =~& -\partial_-\left(\partial_+ + 4\im\right) +
   \left(\partial_A +\im\frac{x_A}{2}\partial_-\right)
   \left(\partial_A -\im\frac{x_A}{2}\partial_-\right) +\\
   &\left(\partial_I +\im \frac{x_I}{2}\partial_-\right)
     \left(\partial_I -\im\frac{x_I}{2}\partial_-\right)-
     \im\Pi^{\ad\bd}\ss_\ad\bar\ss_\bd.
\end{align}

We will be able make the identification
\begin{align}
 U(x^-,x^+,x^A,x^I,\theta^\ad,\bar\theta^\ad,\lambda^a,\bar\lambda^a)
  \leftrightarrow \ket{E_-},
\end{align}
if $U$ satisfy the vacuum conditions described in section
\ref{sec:casimirSpec}. However, in order to have a normalizable state,
the choice of creation and annihilation operators
depend on the sign of $E_-$. The complex linear combinations inside
$\mathfrak{C}_2$ are precisely the ones that appear in
\begin{align}
  &\sp_A +\im \bar\sp_A =
    e^{-\im x^+}\left(-\im\partial_A -\im\frac{1}{2}
    x_AE_-\right),\quad
    \sp_A -\im \bar\sp_A =
    e^{\im x^+}\left(-\im\partial_A +\im\frac{1}{2} x_AE_-\right)
  \\
&\sp_I +\im \bar\sp_I =
    e^{-\im x^+}\left(-\im\partial_I -\im\frac{1}{2}
    x_IE_-\right),\quad
    \sp_I -\im \bar\sp_I =
    e^{\im x^+}\left(-\im\partial_I +\im\frac{1}{2} x_IE_-\right),
\label{creationannahilationOPS}
\end{align}
where we are dropping dependence on $(\bar\sP_A,\bar\sP_I)$ and,
when acting on $U_0$, $\partial_-$ can be identified with $\im
E_-$. The operators above exactly like the creation-annihilation
operators of harmonic oscillators. So we add the dependence on the bosonic
coordinates as the wave function of the ground state of an harmonic
oscillator. This is not surprising since in the light-cone GS
description of the superstring the bosonic directions are massive
world-sheet fields.

From now on we will assume $E_-$ is positive. This means $U$ will be
normalizable if it is annihilated by
\begin{align}\label{BPS1vertex}
  (\sp_A +\im \bar\sp_A)U=(\sp_I +\im \bar\sp_I)U=0.
\end{align}
Up to now the unintegrated vertex operator is
\begin{align}
  U(x^-,x^A,x^I,\theta^\ad,\bar\theta^\ad,\lambda^a,\bar\lambda^a)
  = e^{-\frac{E_-}{4} x^2}
  U_0(x^-,\theta^\ad,\bar\theta^\ad,\lambda^a,\bar\lambda^a),
\end{align}
where $x^2= x^Ax^A+x^Ix^I$.

To get the final constraint on superspace and fix the $x^+$ dependence
we need to define the Clifford vacuum, as discussed in detail in
\cite{Metsaev:2002re}. First we write the quadratic Casimir as
\begin{align}
  \mathfrak{C}_2 =&~~ \partial_-\partial_+ +
   \left(\partial_A +\im\frac{x_A}{2}\partial_-\right)
   \left(\partial_A -\im\frac{x_A}{2}\partial_-\right) +\cr
   &\left(\partial_I +\im \frac{x_I}{2}\partial_-\right)
     \left(\partial_I -\im\frac{x_I}{2}\partial_-\right)+
     \frac{1}{2}\delta_{\ad\cd}\left(\ss_\ad +\im\Pi_{\ad\bd}\bar\ss_\bd\right)
     \left(\ss_\cd -\im\Pi_{\cd\dot{d}}\bar\ss_{\dot{d}}\right).
\end{align}
If we choose the vertex operator $U$ to be annihilated by
$\ss_\ad -\im\Pi_{\ad\bd}\bar\ss_{\bd}$ its $\se_+$ eigenvalue will
be $0$. This choice is the same from the one made in Section
\ref{sec:casimirSpec}. We can study this condition using a chiral basis.
First we define
\begin{align}
  \kappa^\ad =\frac{1}{2}( \theta^\ad +\im\Pi^{\ad\bd}\bar\theta^\bd),\quad
  \bar\kappa^\ad =\frac{1}{2} (\theta^\ad -\im\Pi^{\ad\bd}\bar\theta^\bd),
\end{align}
Next we define a chiral variable
\begin{align}
   \tilde x^- = x^- +\im\kappa^\ad\bar\kappa^\ad.
 \end{align}
The complex combinations of the zero and one $S$-charge supersymmetry generators
$(\ss_\ad,\bar\ss_\ad)$ also have simple expressions in terms of
$(\tilde x^-,\kappa^\ad,\bar\kappa^\ad)$
\begin{align}
  & \ss_\ad +\im\Pi_{\ad\bd}\bar\ss_\bd =
  -\im e^{-\im x^+}\left(\frac{\partial}{\partial\bar\kappa^\ad} -
  2\im\kappa_\ad\tilde\partial_-\right) -
  \im (\theta\sigma^i +\im\bar\theta\sigma_i\Pi)_\ad\left( \sp^{(0)}_i
  +\im\bar\sp^{(0)}_i\right) +\cdots ,\\
  & \ss_\ad -\im\Pi_{\ad\bd}\bar\ss_\bd =
    -\im e^{\im  x^+}\frac{\partial}{\partial\kappa^\ad}
    -\im (\theta\sigma^i -\im\bar\theta\sigma_i\Pi)_\ad
    \left( \sp^{(0)}_i-\im\bar\sp^{(0)}_i\right)+ \cdots
\end{align}

The condition on $U$ can be written in terms of supersymmetry
generators as
\begin{align}\label{BPS2vertex}
  \left( \ss_\ad -\im\Pi_{\ad\bd}\bar\ss_\ad\right)
  U_{\rm final}(\tilde
  x^-,x^+,x^A,x^I,\theta^\ad,\bar\theta^\ad,\theta^a,\bar\theta^a,
  \lambda^a,\bar\lambda^a)=0.
\end{align}
If we restrict to the zero $S$-charge part of $ \ss_\ad
-\im\Pi_{\ad\bd}\bar\ss_\bd$ there would be no solution to this
condition that is compatible with BRST invariance.
However, if we include the higher $(\theta^a,\bar\theta^a)$
contributions in (\ref{RSS}) and (\ref{RSSbar}) we can solve
iteratively in powers of $(\theta^a,\bar\theta^a)$.

Using all the expressions above, we can see that the
vertex operator satisfying the BPS condition (\ref{BPS2vertex}) is
\begin{align}\label{chiralVertex1}
  &U_{\rm final}(x^-,x^A,x^I,\theta^\ad,\bar\theta^\ad,\theta^a,
    \bar\theta^a,\lambda^a,\bar\lambda^a)
  = \\
&~~~~~~
  U^{(0)}(x^-,x^A,x^I,\theta^\ad,\bar\theta^\ad,\lambda^a,\bar\lambda^a)+
  \theta^aU^{(1)}_a + \bar\theta^a \bar U^{(1)}_a +\cdots,
\end{align}
where
\begin{align}
&  U^{(0)}(x^-,x^A,x^I,\theta^\ad,\bar\theta^\ad,\lambda^a,\bar\lambda^a)=\\
& ~~~~~~~~~~~~~~  \lambda^a\bar\lambda^a f_a(E_-,\theta^\ad,0,0,\eta_a)
  \bar f_a(E_-,\bar\theta^\ad,0,0,\bar\eta_a)
  e^{\im E_- \tilde x^-  -\frac{E_-}{4}x^2}
  \Big|_{\eta_a\bar\eta_b=(E_-)^2\Pi_{ab}}
\end{align}
and $(U^{(1)}_a,\bar U^{(1)}_a) $ satisfy
\begin{align}
 & \frac{\partial}{\partial\kappa^\ad}U^{(1)}_a =
  -\im E_-\sigma^i_{a\ad}x_i
  U^{(0)}(x^-,x^A,x^I,\theta^\ad,\bar\theta^\ad,\lambda^a,\bar\lambda^a),\\
 & \frac{\partial}{\partial\kappa^\ad}\bar U^{(1)}_a =
  +\im E_-(\sigma^i\Pi)_{a\ad}x_i
  U^{(0)}(x^-,x^A,x^I,\theta^\ad,\bar\theta^\ad,\lambda^a,\bar\lambda^a).
\end{align}
 Note that $U_{\rm final}$ has a well defined
$\se_+$ charge because of (\ref{fulleplus}) and (\ref{creationannahilationOPS}).

The super
partners of $U$ can organized using
a complex supersymmetry generator
\begin{align}
 \tilde\sq_\ad=  \ss_\ad +\im\Pi_{\ad\bd}\bar\ss_\bd ,
\end{align}
and acting with $\tilde\sq_\ad$ on $U_{\rm final}$ we
generate the whole massless spectrum as described in
\pageref{sec:casimirSpec}.
The spectrum is then
\begin{align}
  &U(\phi)= \phi U_{\rm
  final}(x^-,x^+,x^A,x^I,\theta^\ad,\bar\theta^\ad,\lambda^a,\bar\lambda^a),
\\
  &U(\phi^\ad)=\phi^\ad\tsq_\ad U_{\rm
    final}(x^-,x^+,x^A,x^I,\theta^\ad,\bar\theta^\ad,\lambda^a,\bar\lambda^a)\\
  &U(\phi^{\ad\bd})=\phi^{\ad\bd}\tsq_\bd\tsq_\ad U_{\rm
    final}(x^-,x^+,x^A,x^I,\theta^\ad,\bar\theta^\ad,\lambda^a,\bar\lambda^a)\\
  &U(\phi^{\ad\bd\dot{c}})=\phi^{\ad\bd\dot{c}}\tsq_{\dot{c}}\tsq_\bd\tsq_\ad U_{\rm
    final}(x^-,x^+,x^A,x^I,\theta^\ad,\bar\theta^\ad,\lambda^a,\bar\lambda^a)\\
  &U(\phi^{\ad\bd\dot{c}\dot{d}})=
    \phi^{\ad\bd\dot{c}\dot{d}}\tsq_{\dot{d}}\tsq_{\dot{c}}\tsq_\bd\tsq_\ad
    U_{\rm
    final}(x^-,x^+,x^A,x^I,\theta^\ad,\bar\theta^\ad,\lambda^a,\bar\lambda^a)\\
  &U(\tilde\phi_{\ad\bd\dot{c}})=
   \tilde\phi_{\ad\bd\dot{c}}
   \epsilon^{\ad\bd\dot{c}\ad_1\ad_2\ad_3\ad_4\ad_5}
   \tsq_{\ad_5}\tsq_{\ad_4}\tsq_{\ad_3}\tsq_{\ad_2}\tsq_{\ad_1}
    U_{\rm
    final}(x^-,x^+,x^A,x^I,\theta^\ad,\bar\theta^\ad,\lambda^a,\bar\lambda^a)\\
  &U(\tilde\phi_{\ad\bd})=
   \tilde\phi_{\ad\bd}
   \epsilon^{\ad\bd\ad_1\ad_2\ad_3\ad_4\ad_5\ad_6}
   \tsq_{\ad_6}\tsq_{\ad_5}\tsq_{\ad_4}\tsq_{\ad_3}\tsq_{\ad_2}\tsq_{\ad_1}
    U_{\rm
    final}(x^-,x^+,x^A,x^I,\theta^\ad,\bar\theta^\ad,\lambda^a,\bar\lambda^a)\\
  &U(\tilde\phi_{\ad})=
   \tilde\phi_{\ad}
   \epsilon^{\ad\ad_1\ad_2\ad_3\ad_4\ad_5\ad_6\ad_7}
   \tsq_{\ad_7}\tsq_{\ad_6}\tsq_{\ad_5}\tsq_{\ad_4}\tsq_{\ad_3}\tsq_{\ad_2}\tsq_{\ad_1}
    U_{\rm
    final}(x^-,x^+,x^A,x^I,\theta^\ad,\bar\theta^\ad,\lambda^a,\bar\lambda^a)\\
  &U(\tilde\phi)=
   \tilde\phi
   \epsilon^{\ad_1\ad_2\ad_3\ad_4\ad_5\ad_6\ad_7\ad_8}
   \tsq_{\ad_8}\tsq_{\ad_7}\tsq_{\ad_6}\tsq_{\ad_5}\tsq_{\ad_4}\tsq_{\ad_3}\tsq_{\ad_2}\tsq_{\ad_1}
    U_{\rm final}(x^-,x^+,x^A,x^I,\theta^\ad,\bar\theta^\ad,\lambda^a,\bar\lambda^a).
\end{align}

The set of polarizations
$\{\phi,\phi^\ad,\phi^{\ad\bd},\phi^{\ad\bd\cd},\phi^{\ad\bd\dot{c}\dot{d}},
\tilde\phi_{\ad\bd\dot{c}},\tilde\phi_{\ad\bd},\tilde\phi_\ad,\tilde\phi\}$
describe the 256 supergravity states. Notice that
$\phi^{\ad\bd\dot{c}\dot{d}}$ is self-dual. The vertex operator $ U_{\rm
  final}(x^-,x^+,x^A,x^I,\theta^\ad,\bar\theta^\ad,\lambda^a,\bar\lambda^a)$
is the generating operator for whole massless spectrum. The value of
$E_+$ can be raised further using the bosonic creation operators. This
will add extra dependence on the bosonic coordinates in the form of
Hermite polynomials.

\section{Integrated vertex operator}
\label{sec:IVO}

We will start this section with the most general unintegrated
vertex operator $U=\l^\a \lb^{\b} A_{\a\b}$ satisfying $QU=0$ which implies
\begin{align}
\nabla_{(\a} A_{\b)\bar\g}=\im\g^{\ua}_{\a\b} A_{\ua\g},\quad \bar\nabla_{(\a} A_{\g\b)} = \im\g^{\ua}_{\a\b}A_{\g\ua} .
\label{DA}
\end{align}
The invariance implied by $\d U=Q\L$ gives
\begin{align}
\d A_{\a\bar\b}=\nabla_\a \bar\L_{\b}+\bar\nabla_{\b}\L_\a ,
\label{dA}
\end{align}
where the gauge parameters satisfy
\begin{align}
\nabla_{(\a}\L_{\b)}=\im\g^{\ua}_{\a\b}\L_{\ua},\quad \bar\nabla_{(\a}\bar\L_{\b)}=\im\g^{\ua}_{\a\b}\bar\L_{\ua} .
\label{DL}
\end{align}
In our case, we use light-cone coordinates. It turns out that the physical degrees of freedom in $U$ are in $A_{ab}$, then the other components have to be gauge fixed using (\ref{dA}). Let us check this. Consider the equation for $A_{\dot a \bar\g}$
\begin{align}
\nabla_{(\dot a} A_{\dot b)\bar\g}=\im\d_{\dot a\dot b} A_{-\bar\g} .
\label{eq1}
\end{align}
If there exist a solution of this equation of the form (\ref{dA}), then we can use this gauge symmetry to put $A_{\dot a\bar\g}$ to zero. Try $A_{\dot a\bar\g}=\nabla_{\dot a}\bar\L_{\bar\g}+\nabla_{\bar\g}\L_{\dot a}$ and find the gauge parameters that solve (\ref{eq1}). We obtain
\begin{align}
\nabla_{(\dot a} A_{\dot b)\bar\g}=\{\nabla_{\dot a},\nabla_{\dot b}\}+\nabla_{(\dot a}\nabla_{\bar\g}\L_{\dot b)}=-T_{\dot a\dot b}{}^-\nabla_-\bar\L_{\bar\g}-\nabla_{\bar\g}\nabla_{(\dot a}\L_{\dot b)}+\{\nabla_{\bar\g},\nabla_{(\dot a}\}\L_{\dot b)} .
\end{align}
Using the values of torsion and (\ref{DL})
\begin{align}
\nabla_{(\dot a} A_{\dot b)\bar\g}=\im\d_{\dot a\dot b} ( \nabla_-\bar\L_{\bar\g}-\nabla_{\bar\g}\L_-) +\{\nabla_{\bar\g},\nabla_{(\dot a}\}\L_{\dot b)} .
\end{align}
The last anti-commutator is a curvature. It vanishes for $\bar\g=\dot{\bar c}$. For $\bar\g=\bar c$, it is proportional to $(\s^i\Pi)_{(\dot a\bar c}\s^i_{d\dot b)} \L_d$ which is proportional to $\d_{\dot a\dot b}	\Pi_{\bar c d}\L_d$. In summary, the equation (\ref{eq1}) is satisfied by a gauge transformation expression. Then, $A_{\dot a\bar\g}$ can be gauge fixed to zero. Similarly, $A_{\g \dot{\bar a}}$ can be put to zero.

In the gauge $A_{\ad\bar\g}=A_{\g\dot{\bar a}}=0$, the unintegrated vertex operator is $U=\l^a\lb^b A_{ab}$ and the superfield $A_{ab}$ satisfies the equations
\begin{align}
\nabla_{(a}A_{b)c}=\im\d_{ab}A_{+c},\quad \nabla_{\dot a}A_{bc}=\im\s^i_{b\dot a}A_{ic},\quad \bar\nabla_{(a} A_{cb)}=\im\d_{ab} A_{c+},\quad \bar\nabla_{\ad}A_{cb}=\im\s^i_{b\ad} A_{ci} .
\label{DAf}
\end{align}
From here, the second fermionic covariant derivatives of $A_{ab}$ are constrained to satisfy
\begin{align}
\nonumber
&\nabla_a A_{+c}-\nabla_+ A_{ac}=0,\quad \nabla_- A_{ac}=2W_{ac},\quad \nabla_a A_{ic}-\nabla_i A_{ac}=\s^i_{a\dot b} W_{\dot b c},\\& \nonumber \nabla_{\dot a}A_{+c}=-2W_{\dot a c},\quad \nabla_{\dot a} A_{ic}=\s^i_{b\dot a} W_{bc} ,\\ &\nonumber \bar\nabla_a A_{c+}-\nabla_+ A_{ca}=0,\quad \nabla_- A_{c a}=2\bar W_{ca},\quad \bar\nabla_a A_{ci}-\nabla_i A_{ca}=\s^i_{a\dot{a}}\bar W_{c\dot{b}},\\ &\bar\nabla_{\dot{a}} A_{c+}=-2\bar W_{c\dot{a}},\quad \bar\nabla_{\dot{a}}A_{ci}=\s^i_{b\dot{a}} \bar W_{cb} .
\label{DAW}
\end{align}
Note that $W_{ab}=\bar W_{ab}$.
The next group of equations come from performing the third fermionic covariant derivative of $A_{ab}$. These equations depend on the field strengths $F_{\ua\ub\g}=\nabla_{[\ua}A_{\ub]\g}$ and $F_{\g\ua\ub}=\nabla_{[\ua}A_{\g\ub]}$ and they are
\begin{align}
\nonumber
& \nabla_a W_{bc}=-\frac{\im}{4}\d_{ab}F_{+-c}+\frac{\im}{4}(\s_{ij})_{ab}F_{ijc},\quad \nabla_{\dot a} W_{bc}=\frac{\im}{2}\s^i_{b\dot a}F_{-ic} ,\\& \nonumber \nabla_a W_{\dot b c}-\frac12\Pi_{\dot b\dot{c}}\bar\nabla_{\dot{c}} A_{ac}=\frac{\im}{2}\s^i_{a\dot b} F_{+ic},\quad \nabla_{\dot a}W_{\dot b c}=\frac{\im}{4}\d_{\dot a\dot b} F_{+-c}+\frac{\im}{4}(\s_{ij})_{\dot a\dot b} F_{ijc} ,\\& \nonumber \bar\nabla_{a} W_{cb} =-\frac{\im}{4} \d_{ab} F_{c+-}+\frac{\im}{4}(\s_{ij})_{ab} F_{cij},\quad \bar\nabla_{\dot{a}} W_{cb}=\frac{\im}{2}\s^i_{b\dot{a}}F_{c-i} ,\\& \bar\nabla_{a}\bar W_{c\dot{b}}+\frac12\Pi_{\dot{b}\cd}\nabla_{\dot c} A_{ca}=\frac{\im}{2}\s^i_{a\dot{b}}F_{c+i},\quad \bar\nabla_{\dot{a}}\bar W_{c\dot{b}}=\frac{\im}{4}\d_{\dot{a}\dot{b}} F_{c+-}+\frac{\im}{4}(\s_{ij})_{\dot{a}\dot{b}} F_{cij} .
\label{DWF}
\end{align}

As in flat space, the integrated vertex operator ${\cal V}$ is obtained from the equations
\begin{align}
Q{\cal V}=\p \bar{\cal W} -\bar\p{\cal W},\quad Q\bar{\cal W}=\bar\p U,\quad Q{\cal W}=\p U .
\label{defV}
\end{align}
Note that we can use the equations of motion derived from the action (\ref{action}). The equations we need are
\begin{align}
\bar J_1^a=J_3^a=\bar\nabla\l^a=\nabla\bar\l^a=0,\quad \bar J_1^\ad=-\frac{\im}{2}\Pi^{\ad\bd}\bar d_\bd,\quad J_3^\ad=\frac{\im}{2}\Pi^{\ad\bd}d_\bd .
\label{eoms}
\end{align}
The ${\cal W}$  satisfying $Q{\cal W}=\p U$ is given by
\begin{align}
\nonumber {\cal W}=&~~\lb^b(J_1^a A_{ab}+ J^+ A_{+b} +J^i A_{ib} -\im
                     d_a W_{ab} - \im d_{\dot a} W_{\dot a b} \\& +
                    N^{+-}F_{+-b}+N^{+i}F_{+ib}+N^{-i}F_{-ib}+\frac{1}{2}
                   N^{ij} F_{ij b} ) \equiv \lb^b ( J_1^a A_{ab} + \varphi_b ) ,
\label{wpu}
\end{align}
where $\varphi_b$ is defined by (\ref{wpu}) and
\begin{align}
\nonumber &N^{+-}=-\frac14 \l^a\o_a +\frac14\l^\ad\o_\ad,\quad N^{+i}=\frac12\s^i_{a\bd}\l^a\o_\bd,\\ &N^{-i}=\frac12\s^i_{a\bd}\l^\bd\o_a,\quad N^{ij}=\frac12\l^a\o_b\s^{ij}_{ab}+\frac12\l^\ad\o_\bd\s^{ij}_{\ad\bd} .
\label{Ns}
\end{align}
The BRST transformations are $Q\o_a=d_a, Q\o_\ad=d_\ad$ and
\begin{align}
\nonumber &Q J^a=\nabla\l^a,\quad Q J^+=\im(\l^a J_1^a+\lb^a J_3^a),\quad Q J^i=\im\s^i_{a\bd}(\l^{(a} J_1^{\bd)}+\lb^{(a} J_3^{\bd)}),\quad  \\& \nonumber Qd_a=\im(2\l_a J^- -\s^i_{a\bd}\l^\bd J^i) -\frac{\im}{2} (\s_{[ij]}\Pi)_{ab}\lb^b N^{[ij]}+\im(\s_i\Pi)_{a\bd} \lb^\bd N^{+i}, \\& Q d_\ad=\im(2\l^\ad J^+ - \s^i_{b\ad} \l^b J^i) - \im (\s_i\Pi)_{b\ad}\lb^b N^{+i} .
\label{brstL}
\end{align}
Consider the BRST transformation of the first term in (\ref{wpu}). It contains $\lb^b\nabla\l^a A_{ab}$ which is equal to $\nabla(\lb^b\l^a A_{ab})-\lb^b\l^a\nabla A_{ab}=\nabla U-\lb^b\l^a\nabla A_{ab}$. Here we are using the equations of motion (\ref{eoms}). Then
\begin{align}
\nonumber Q{\cal W}=&~~\p U\cr &+ \lb^b ( -\l^a [ J_1^c \nabla_c A_{ab} + J_1^\cd\nabla_\cd A_{ab} + J_3^\cd\bar\nabla_\cd A_{ab}+J^+\nabla_+ A_{ab}+ J^-\nabla_- A_{ab} + J^i A_{ab} ] \\& \nonumber -J_1^a [ \l^c\nabla_c A_{ab}+\l^\cd\nabla_\cd A_{ab}+\lb^c\bar\nabla_c A_{ab}+\lb^\cd \bar\nabla_\cd A_{ab}] +\im [ \l^c J_1^c ] A_{+b} \\& \nonumber +J^+ [ \l^c \nabla_c A_{+b} + \l^\cd \nabla_\cd A_{+b} + \lb^c \bar\nabla_c A_{+b} + \lb^\cd\bar\nabla_\cd A_{+b} ] + \im\s^i_{c\dd} [ \l^c J_1^\dd + \l^\dd J_1^c + \lb^c J_3^\dd ] A_{ib} \\& \nonumber + J^i [ \l^c\nabla_c A_{ib} + \l^\cd \nabla_\cd A_{ib}+ \lb^c \bar\nabla_c A_{ib} +\lb^\cd \bar\nabla_\cd A_{ib} ] \\& \nonumber +[2\l^a J^--\s^i_{a\cd}\l^\cd J^i -\frac{1}{2} (\s_{[ij]}\Pi)_{ac}\lb^c N^{[ij]}+(\s_i\Pi)_{a\cd} \lb^\cd N^{+i}] W_{ab} \\& \nonumber +\im d_a [ \l^c \nabla_c W_{ab} + \l^\cd \nabla_\cd W_{ab} + \lb^c \bar\nabla_c W_{ab} + \lb^\cd \bar \nabla_\cd W_{ab} ] \\& \nonumber +[2\l^\ad J^+ - \s^i_{c\ad} \l^c J^i - (\s_i\Pi)_{c\ad}\lb^c N^{+i} ] W_{\ad b}\\& \nonumber +\im d_\ad [ \l^c \nabla_c W_{\ad b} + \l^\cd \nabla_\cd W_{\ad b} + \lb^c \bar\nabla_c W_{\ad b} + \lb^\cd�\bar\nabla_\cd W_{\ad b} ] +\frac14 [ \l^a d_b \s^{ij}_{ab} + \l^\ad d_\bd \s^{ij}_{\ad\bd}] F_{ijb}\\& \nonumber +\frac12 N^{ij} [ \l^c \nabla_c F_{ijb} + \l^\cd \nabla_\cd F_{ijb} +\lb^c\bar\nabla_c F_{ijb} +\lb^\cd\bar\nabla_\cd F_{ijb}] +\frac14 [ -\l^a d_a + \l^\ad d_\ad ] F_{+-b}\\& \nonumber +N^{+-} [ \l^c \nabla_c F_{+-b} + \l^\cd \nabla_\cd F_{+-b} + \lb^c\bar\nabla_c F_{+-b} + \lb^\cd \bar\nabla_\cd F_{+-b} ] + \frac12 [ \s^i_{a\bd} \l^a d_\bd ] F_{+ib}\\& \nonumber N^{+i} [ \l^c \nabla_c F_{+ib} + \l^\cd \nabla_\cd F_{+ib} + \lb^c \bar\nabla_c F_{+ib} + \lb^\cd \bar\nabla_\cd F_{+ib} ] + \frac12 [�\s^i_{a\bd} \l^\bd d_a ] F_{-ib} \\& +N^{-i} [ \l^c \nabla_c F_{-ib} + \l^\cd \nabla_\cd F_{-ib} + \lb^c \bar\nabla_c F_{-ib} + \lb^\cd \bar\nabla_\cd F_{-ib} ] ) .
\label{remains}
\end{align}
Here we used the equation of motion $J_3^a=0$.  It is easy to check that the terms with $\l^\a J_1^\b$ and are zero because the equations (\ref{DAf}). The term with $\l^a J_3^\ad$ mixes with terms with $\l^a d_\ad$, because the last equation in (\ref{eoms}), and the result is one of the equations in (\ref{DWF}). The terms with $\l^\a J_2$ also vanish because the equations (\ref{DAW}). Similarly, the terms involving $\l^\a d_\b$ also vanish. They factor the equations (\ref{DWF}). Up to now, we are left with terms with $\o_\a$ and terms quadratic in $\lb$. They are
 \begin{align}
\nonumber& \lb^b ( -J_1^a [ \lb^c\bar\nabla_c A_{ab}+\lb^\cd \bar\nabla_\cd A_{ab}] + J^+ [ \lb^c \bar\nabla_c A_{+b} + \lb^\cd\bar\nabla_\cd A_{+b} ] + \im\s^i_{c\dd} [ \lb^c J_3^\dd ] A_{ib} \\ \nonumber &+ J^i [ \lb^c \bar\nabla_c A_{ib} +\lb^\cd \bar\nabla_\cd A_{ib} ] + [ -\frac{1}{2} (\s_{[ij]}\Pi)_{ac}\lb^c N^{[ij]}+(\s_i\Pi)_{a\cd} \lb^\cd N^{+i}] W_{ab} \\ \nonumber &+\im d_a [ \lb^c \bar\nabla_c W_{ab} + \lb^\cd \bar \nabla_\cd W_{ab} ]  + [ - (\s_i\Pi)_{c\ad}\lb^c N^{+i} ] W_{\ad b} +\im d_\ad [ \lb^c \bar\nabla_c W_{\ad b} + \lb^\cd�\bar\nabla_\cd W_{\ad b} ] \\ \nonumber &+ \frac12 N^{ij} [ \l^c \nabla_c F_{ijb} + \l^\cd \nabla_\cd F_{ijb} +\lb^c\bar\nabla_c F_{ijb} +\lb^\cd\bar\nabla_\cd F_{ijb}] \\ \nonumber &+N^{+-} [ \l^c \nabla_c F_{+-b} + \l^\cd \nabla_\cd F_{+-b} + \lb^c\bar\nabla_c F_{+-b} + \lb^\cd \bar\nabla_\cd F_{+-b} ] \\ \nonumber &+N^{+i} [ \l^c \nabla_c F_{+ib} + \l^\cd \nabla_\cd F_{+ib} + \lb^c \bar\nabla_c F_{+ib} + \lb^\cd \bar\nabla_\cd F_{+ib} ]  \\ &+N^{-i} [ \l^c \nabla_c F_{-ib} + \l^\cd \nabla_\cd F_{-ib} + \lb^c \bar\nabla_c F_{-ib} + \lb^\cd \bar\nabla_\cd F_{-ib} ] ) .
\label{remains2}
\end{align}
The terms quadratic in $\lb$ will vanish because of the pure spinor conditions $\lb^a\lb^a=\s^i_{a\bd}\lb^a\lb^\bd=0$ and the commutation relations $[\nabla_A,\nabla_B]$ which are given by the torsion and curvature of the plane wave background. These kind of terms with $J_1^a$ in (\ref{remains2}) zero because of (\ref{DAf}). There are two terms with $J^+$, the first goes with
\begin{align}& \lb^b \lb^c \bar\nabla_c A_{+b}=-\frac{\im}{8}\lb^b
  \lb^c \bar\nabla_c \nabla_d A_{db}=\frac{\im}{8}\lb^b \lb^c \nabla_d
  \bar\nabla_c A_{db} =0\\
  & \frac{\im}{16}\lb^b \lb^c \nabla_d \bar\nabla_{(c} A_{db)}
 =-\frac{1}{16}\lb^b \lb^b \nabla_d A_{d+}=0.\label{lb0} \end{align}
Similarly, the second term with $J^+$ and the terms with $J^i$ also vanish. The last term in the first line of (\ref{remains2}) combines with the term involving $\lb^c$ in the last term of the third line of (\ref{remains2}). They go with
\begin{align} \lb^b\lb^c ( \bar\nabla_c W_{\ad b} + \frac{\im}{2} (\s^i\Pi)_{c\ad} A_{ib} ) ,\label{lb1} \end{align}
but
\begin{align}
\bar\nabla_c W_{\ad b}=-\frac12\bar\nabla_c\nabla_\ad
  A_{+b}=-\frac12\{\nabla_\ad,\bar\nabla_c\}A_{+b}+\frac12\nabla_\ad\bar\nabla_c
  A_{+b} .
\end{align}
Note that the last term will vanish in (\ref{lb1}) because of (\ref{lb0}). Using the non-vanishing curvature component for the commutator are $R_{\ad c+}{}^i = -\im(\s^i\Pi)_{c\ad}$ and $R_{\ad cb}{}^\dd=-\frac{\im}{2}\s^i_{b\dd}(\s^i\Pi)_{c\ad}$ we obtain that
\begin{align}
\bar\nabla_c W_{\ad b}=-\frac{\im}{2} (\s^i\Pi)_{c\ad} A_{ib}
  -\frac{\im}{4}\s^i_{b\dd}(\s^i\Pi)_{c\ad} A_{+\dd},
\end{align}
but the last term vanishes in (\ref{lb1}) because $\lb^b\lb^c \s^i_{b\cd} \s^i_{c\dd}=0$. The first term here cancel the second term in (\ref{lb1}). Then, we have verified that (\ref{lb1}) is zero. The terms with $d_\a$ in (\ref{remains2}) are also zero because (\ref{DWF}) and pure spinor conditions. Up to this point we have terms with $\o_\a$, which are
\begin{align}
\nonumber& \lb^b (  [ -\frac{1}{2} (\s_{[ij]}\Pi)_{ac}\lb^c
           N^{[ij]}+(\s_i\Pi)_{a\cd} \lb^\cd N^{+i}] W_{ab}  + [ -
           (\s_i\Pi)_{c\ad}\lb^c N^{+i} ] W_{\ad b} \\ \nonumber & + \frac12 N^{ij} [ \l^c \nabla_c F_{ijb} + \l^\cd \nabla_\cd F_{ijb} +\lb^c\bar\nabla_c F_{ijb} +\lb^\cd\bar\nabla_\cd F_{ijb}] \\ \nonumber &+N^{+-} [ \l^c \nabla_c F_{+-b} + \l^\cd \nabla_\cd F_{+-b} + \lb^c\bar\nabla_c F_{+-b} + \lb^\cd \bar\nabla_\cd F_{+-b} ] \\ \nonumber &+N^{+i} [ \l^c \nabla_c F_{+ib} + \l^\cd \nabla_\cd F_{+ib} + \lb^c \bar\nabla_c F_{+ib} + \lb^\cd \bar\nabla_\cd F_{+ib} ]  \\ &+N^{-i} [ \l^c \nabla_c F_{-ib} + \l^\cd \nabla_\cd F_{-ib} + \lb^c \bar\nabla_c F_{-ib} + \lb^\cd \bar\nabla_\cd F_{-ib} ] ) .
\label{remains3}
\end{align}
The terms involving $\l^\a$ have the form $N^{\ua\ub}\l^\a \nabla_\a F_{\ua\ub\b}$. Because of the pure spinor condition, this expression becomes proportional to the equation of motion for $W^\a{}_\b$ (see section 4 of \cite{Chandia:2017afc}). That remains are quadratic in $\lb$, the part with $F$ had the form $\lb^\b\lb^\a \bar\nabla_\a F_{\ua\ub\b}$. This expression combines with the first line in (\ref{remains3}) to vanish (see again section 4 of \cite{Chandia:2017afc}). A similar calculation gives $\bar{\cal W}$, satisfying $Q\bar{\cal W}=\bar\p U$, given by
\begin{align}
\bar{\cal W}=&~~ \nonumber \l^a ( \bar J_3^b A_{ab} + \bar J^+ A_{a+} + \bar J^i A_{ai} - \im \bar d_b W_{ab} - \im \bar d_\bd \bar W_{a\bd} \\ &+ \bar N^{+-}F_{a+-}+\bar N^{+i}F_{a+i}+\bar N^{-i}F_{a-i}+\frac{1}{2} \bar N^{ij} F_{aij} ) \equiv \l^a ( \bar J_3^b A_{ab} + \bar\varphi_a) ,
\label{wpub}
\end{align}
where
\begin{align}
\nonumber &\bar N^{+-}=-\frac14 \lb^a\ob_a +\frac14\lb^\ad\ob_\ad,\quad \bar
            N^{+i}=\frac12\s^i_{a\bd}\lb^a\ob_\bd,\\ &
          \bar N^{-i}=\frac12\s^i_{a\bd}\lb^\bd\ob_a,\quad
    \bar N^{ij}=\frac12\lb^a\ob_b\s^{ij}_{ab}+\frac12\lb^\ad\ob_\bd\s^{ij}_{\ad\bd} .
\label{Nsb}
\end{align}

A further simplification appears if we use the $x$-dependence of the
superfields in the form $e^{\im E_-  x^- - \frac{E_-}{4} x^2}$. We
have that
\begin{align}
\nonumber {\cal W}=&~~\lb^b \Big( [ J_1^a +\frac{E_-}{2} d_a - \frac{E_-}{4}
  d_\ad \s^i_{a\ad} x^i ] A_{ab} + [ J^+ -\im E_- N^{+-} +
  \frac{E_-}{2} N^{+i} x^i ] A_{+b} \\ \nonumber &+ [ J^i + \im E_-
  N^{-i} + \frac{E_-}{2} N^{ij} x^j ] A_{ib} + N^{+i} \nabla_+ A_{ib}
  \Big)  ,\\ \nonumber \bar{\cal W}=&~~\l^a ( [ \bar J_3^b + \frac{E_-}{2} \bar d_b - \frac{E_-}{4} \bar d_\bd  \s^i_{b\bd} x^i  ] A_{ab} + [ \bar J^+ -\im E_- \bar N^{+-} + \frac{E_-}{4} \bar N^{+i} x^i ] A_{a+} \\ &+ [ \bar J^i + \im E_- \bar N^{-i} +\frac{E_-}{2} \bar N^{ij} x^j ]A_{ai} + \bar N^{+i} \nabla_+ A_{ai} ) .
\label{Wsimp}
\end{align}

The integrated vertex ${\cal V}$ is computed,
as in flat space, by satisfying (\ref{defV}). In order to write a
shorter expression we use in some terms of ${\cal V}$ a
ten-dimensional covariant notation with a vector index $\ua=(+, -, i)$
and a spinor index $\a=(a,\ad)$. The integrated is given by
\begin{align}
\nonumber {\cal V}= J^a_1 \bar J^b_3 A_{ab} + J^a_1\bar\varphi_a - \bar J^a_3\varphi_a  + J^\ua \bar J^\ub A_{\uab}  + d_\a \bar J^\ua E_\ua{}^\a + J^\ua \bar d_\a \bar E_\ua{}^\a + J^\uc \bar N^{\ua\ub} \O_{\uc\ua\ub}  \\ + N^{\ua\ub} \bar J^\uc \bar\O_{\uc\ua\ub} + d_\a \bar d_\b P^{\a\b} + d_\a \bar N^{\ua\ub} C_{\ua\ub}{}^\a + N^{\ua\ub} \bar d_\a \bar C_{\ua\ub}{}^\a + N^{\ua\ub} \bar N^{\uc\ud} R_{\ua\ub\uc\ud} ,
\label{IVO}
\end{align}
where $\varphi_a$ is defined in (\ref{wpu}) and $\bar\varphi_a$
is defined in (\ref{wpub}). The following is a summary of the results.
After imposing (\ref{defV}), the non-zero components of $A$ satisfies
\begin{align}
\nonumber &\nabla_{(a} A_{b)+} = \im \d_{ab} A_{++},\quad \bar\nabla_{(a} A_{+b)} = -\im \d_{ab} A_{++},\quad \nabla_{(a} A_{b)i} = \im \d_{ab} A_{+i},\quad \bar\nabla_{\ad} A_{+a} = -\im \s^i_{a\ad}  A_{+i}, \\ &\bar\nabla_{(a} A_{i b)} = -\im \d_{ab} A_{i+},\quad \nabla_{\ad} A_{a+} = \im \s^i_{a\ad} A_{i+},\quad \nabla_\ad A_{ai} = \im \s^j_{a\ad} A_{ji},\quad \bar\nabla_\ad A_{ia} = -\im \s^j_{a\ad} A_{ij} .
\label{A}
\end{align}
The non-zero components of $E$ satisfy
\begin{align}
E_{+a}=-\frac{\im}{2}\nabla_- A_{a+},\quad E_{ia}=-\frac{\im}{2}\nabla_- A_{ai} .
\label{E}
\end{align}
The non-zero components of $\bar E$ satisfy
\begin{align}
\bar E_{+a} = \frac{\im}{2} \nabla_- A_{+a},\quad \bar E_{ia} = \frac{\im}{2} \nabla_- A_{ia},\quad \bar E_{+\ad}=\frac{\im}{2}\bar\nabla_\ad A_{++},\quad \bar E_{i\ad} = \frac{\im}{2} \bar\nabla_\ad A_{i+}
\label{Eb}
\end{align}
The non-zero components of $\O$ are
\begin{align}
\nonumber &\O_{++-}=-\frac12\bar\nabla_a \bar E_{+a},\quad \O_{i+-}=\frac12\bar\nabla_a \bar E_{ia},\quad \O_{++i}=\frac14 (\s_i)_{a\ad} \bar\nabla_a \bar E_{+\ad} + \frac{\im}{8} (\s_i\Pi)_{a\ad} \nabla_\ad A_{+a},\\ \nonumber &\O_{j+i}=\frac14 (\s_i)_{a\ad} \bar\nabla_a \bar E_{i\ad} + \frac{\im}{8} (\s_j\Pi)_{a\ad} \nabla_\ad A_{ia},\quad \O_{+-i} = \frac14 (\s_i)_{a\ad} \bar \nabla_\ad  \bar E_{+a} ,\quad \O_{j-i} = \frac14 (\s_i)_{a\ad} \bar\nabla_\ad \bar E_{ja} \\ &\O_{+ij} = \frac18 (\s_{ij})_{ab} \bar\nabla_a \bar E_{+b},\quad \O_{kij} = \frac18 (\s_{ij})_{ab} \bar\nabla_a \bar E_{kb} . ~~~~~~~~~~~~~~~~~
\label{O}
\end{align}
The non-zero components of $\bar\O$ are
\begin{align}
\nonumber & \bar\O_{++-}=-\frac12\nabla_a E_{+a},\quad
  \bar\O_{i+-}=-\frac12\nabla_a E_{ia},\quad
  \bar\O_{++i}=\frac{\im}{8}(\s_i\Pi)_{a\ad}\bar\nabla_\ad A_{a+},\\
  \nonumber  &\bar\O_{j+i}=\frac{\im}{8}(\s_j\Pi)_{a\ad}\bar\nabla_\ad
               A_{ai},\quad  \bar\O_{+-i} = \frac14 (\s_i)_{a\ad}
               \nabla_\ad  E_{+a},\quad \bar\O_{j-i} = \frac14 (\s_i)_{a\ad} \nabla_\ad E_{ja},\\ &\bar\O_{+ij} = \frac18 (\s_{ij})_{ab} \nabla_a E_{+b},\quad \bar\O_{kij} = \frac18 (\s_{ij})_{ab} \nabla_a E_{kb} .
\label{Ob}
\end{align}
The non-zero components of $P$ are
\begin{align}
P_{ab}=\frac12 \nabla_- W_{ab},\quad P_{a\ad}=-\frac{\im}{2} \bar\nabla_\ad E_{+a},\quad P_{\ad a}=-\frac{\im}{2} \nabla_\ad \bar E_{+a},\quad P_{\ad\bd}=-\frac{\im}{2} \nabla_\ad \bar E_{+\bd} .
\label{P}
\end{align}
The non-zero components of $C$ are
\begin{align}
\nonumber & C_{+-a} = -\frac{\im}{2} \nabla_- F_{a+-},\quad C_{+-\ad} =
  \frac{\im}{2} \nabla_\ad \O_{++-},\quad C_{+ia} = -\frac{\im}{2}
  \nabla_- F_{a+i},\\ \nonumber & C_{+i\ad} = \frac12(\s_i\Pi)_{a\ad}
     \bar E_{+a} + \frac{\im}{2} \nabla_\ad \O_{++i},\quad C_{-ia} =
        -\frac{\im}{2} \nabla_- F_{a-i},\quad C_{-i\ad} =
        \frac{\im}{2} \nabla_\ad \O_{+-i},\\
          & C_{ija} = -\frac{\im}{4} \nabla_- F_{aij},\quad
         C_{ij\ad} = \frac{\im}{2} \nabla_\ad \O_{+ij} .
\label{C}
\end{align}
The non-zero components of $\bar C$ are
\begin{align}
\nonumber &\bar C_{+-a} = \frac{\im}{2} \nabla_- F_{+-a},\quad \bar C_{+-\ad} = \frac{\im}{2} \bar\nabla_\ad \O_{++-},\quad \bar C_{+ia} = \frac{\im}{2} \nabla_- F_{+ia} \\ \nonumber &\bar C_{+i\ad} = -\frac12(\s_i\Pi)_{a\ad}  E_{+a} + \frac{\im}{2} \bar\nabla_\ad \bar\O_{++i},\quad \bar C_{-ia} = \frac{\im}{2} \nabla_- F_{-ia},\quad \bar C_{-i\ad} = \frac{\im}{2} \bar\nabla_\ad \bar \O_{+-i} \\ &\bar C_{ija} = \frac{\im}{4} \nabla_- F_{ija},\quad  \bar C_{ij\ad} = \frac{\im}{2} \bar\nabla_\ad \bar \O_{+ij} .
\label{Cb}
\end{align}
Finally, the non.zero components of $R$ are
\begin{align}
\nonumber & R_{+-+-}=-\frac12 \nabla_a C_{+-a},\quad
            R_{+-+i}=-\frac{\im}{2}(\s_i\Pi)_{a\ad} P_{a\ad} - \frac12
            \nabla_a C_{+ia},\\ \nonumber
          &R_{+i+-}=-\frac{\im}{2}(\s_i\Pi)_{a\ad} P_{\ad a} - \frac12
            \bar\nabla_a \bar C_{+ia},\quad R_{+--i}= -\frac12
            \nabla_a C_{-ia},\quad R_{-i+-}= -\frac12 \bar\nabla_a
            \bar C_{-ia},\\ \nonumber
          &R_{+-ij}=\frac{\im}{8}(\s_{[i}\Pi\s_{j]})_{ab} P_{ba} -
            \frac12 \nabla_a C_{ija},\quad R_{ij+-}= \frac{\im}{8}(\s_{[i}\Pi\s_{j]})_{ab} P_{ab} - \frac12 \bar\nabla_a \bar C_{ija},\\ \nonumber&R_{+i+j}=\frac14 (\s_i)_{a\ad} \nabla_a C_{+j\ad} +\frac{\im}{8}(\s_i\Pi)_{a\ad} \bar\nabla_\ad F_{a+j},\\ \nonumber &R_{+i-j}= \frac{\im}{4} (\s_i\Pi\s_j)_{ab} P_{ab} + \frac14 (\s_j)_{a\ad} \bar\nabla_\ad \bar C_{+ia},\quad R_{-j+i}= \frac{\im}{4} (\s_i\Pi\s_j)_{ab} P_{ba} + \frac14 (\s_j)_{a\ad} \nabla_\ad C_{+ia}, \\ \nonumber &R_{+ijk}= -\frac{\im}{8} (\s_{jk}\s_i\Pi)_{a\ad} P_{\ad a} + \frac18 (\s_{jk})_{ab} \bar\nabla_a \bar C_{+ib},\\ \nonumber&R_{jk+i}= -\frac{\im}{8} (\s_{jk}\s_i\Pi)_{a\ad} P_{a\ad} + \frac18 (\s_{jk})_{ab} \nabla_a  C_{+ib}, \\ \nonumber &R_{-i-j}=\frac14 (\s_i)_{a\ad} \nabla_\ad C_{-ja},\quad R_{-ijk}=\frac14 (\s_i)_{a\ad} \nabla_\ad C_{jka},\quad R_{jk-i}=\frac14 (\s_i)_{a\ad} \bar\nabla_\ad \bar C_{jka}, \\ &R_{ijkl} = -\frac{\im}{32} (\s_{[k}\Pi\s_{l]}\s_{ij})_{ab} P_{ba} + \frac18 (\s_{ij})_{ab} \nabla_a C_{klb} .
\label{R}
\end{align}
This completes the calculation of all superfields in the integrated
vertex operator (\ref{IVO}).

\vskip 0.3in
{\bf Acknowledgements}~
We would like to thank William Linch for useful discussions and
comments on the draft. The work of B{\sc cv} is partially
supported by FONDECYT
grant number 1151409 and  CONICYT grant number DPI20140115.

\appendix

\section{Conventions}
\label{app1}

\def\rad{{|\lambda\bar\lambda\rangle}}
\def\ll{{\langle}}
\def\rr{{\rangle}}

\def\a{{\alpha}}
\def\ah{{\widehat\alpha}}
\def\l{{\lambda}}
\def\lh{{\widehat\lambda}}
\def\b{{\beta}}
\def\bh{{\widehat\beta}}
\def\g{{\gamma}}
\def\gh{{\widehat\gamma}}
\def\d{{\delta}}
\def\dh{{\widehat\delta}}
\def\e{{\epsilon}}
\def\ve{{\varepsilon}}
\def\veb{\overline\varepsilon}
\def\s{{\sigma}}
\def\r{{\rho}}
\def\s{{\sigma}}
\def\rh{{\widehat\rho}}
\def\N{{\nabla}}
\def\Nb{{\overline\nabla}}
\def\O{{\Omega}}
\def\Oh{{\widehat\O}}
\def\o{{\omega}}
\def\half{{1\over 2}}
\def\p{{\partial}}
\def\pb{{\overline\partial}}
\def\t{{\theta}}
\def\th{{\widehat\theta}}
\def\ph{{\widehat p}}
\def\oh{{\widehat\o}}
\def\L{{\Lambda}}
\def\Lh{{\widehat\L}}
\def\dhh{{\widehat d}}
\def\zb{{\overline z}}
\def\Qb{{\overline Q}}
\def\yb{{\overline y}}
\def\Pih{{\widehat\Pi}}
\def\Qt{{\widetilde Q}}
\def\Qh{{\widehat Q}}
\def\add{{\dot a}}
\def\bdd{{\dot b}}
\def\cdd{{\dot c}}
\def\ddd{{\dot d}}
\def\Kb{\overline K}
\def\Jb{\overline J}
\def\jb{\overline j}
\def\Ph{\widehat P}
\def\Tb{\overline T}
\def\Ab{\overline A}

We will use an $\mathfrak{so}(4)\oplus \mathfrak{so}(4)$ decomposition of the  $\mathfrak{psu}(2,2|4)$ algebra. The generators will be denoted by:
\begin{align}
\frakg_0=(\sM_{AB}, \sM_A, \sM_{IJ}, \sM_I),\quad \frakg_1=(\sQ_a, \sQ_\add),\quad \frakg_2=(\sT, \sP_A, \sJ, \sP_I),\quad \frakg_3=(\bar\sQ_a, \bar\sQ_\add) .
\end{align}
Translations and supercharges $(\frakg_1,\frakg_2,\frakg_3)$ will
be hermitian and rotations $\frakg_0$ will be
anti-hermitian. This convention removes all $\im$ from the bosonic
commutators. The non zero commutators are
($R$ is the radius of $AdS_5$ and $S^5$):
\begin{align}
&  [\sM_{AB}, \sM_{CD}] = - \d_{A[C} \sM_{D]B} + \d_{B[C} \sM_{D]A} ,\\
&  [\sM_{AB}, \sM_C] = \d_{C[A} \sM_{B]} ,\quad [\sM_A, \sM_B] = -\sM_{AB} ,\\
&[\sM_{IJ}, \sM_{KL}] = - \d_{I[K} \sM_{L]J} + \d_{J[K} \sM_{L]I} ,\quad [\sM_{IJ}, \sM_K] = \d_{K[I} \sM_{J]} ,\quad [\sM_I, \sM_J] = \sM_{IJ},\\
&[\sT,\sP_A] = -\frac{1}{R^2}\sM_A,\quad [\sP_A,\sP_B]= -\frac{1}{R^2}\sM_{AB},\quad [\sJ,\sP_I]= \frac{1}{R^2}\sM_I,\quad [\sP_I,\sP_J]=\frac{1}{R^2} \sM_{IJ},\\
&[\sM_{AB},\sP_C]= \d_{C[A} \sP_{B]},\quad [\sM_A,\sT]= -\sP_A ,\quad [\sM_A,\sP_B]= -\d_{AB} \sT,\\
&[\sM_{IJ},\sP_K]= \d_{K[I} \sP_{J]} ,\quad [\sM_I,\sJ]=  \sP_I,\quad [\sM_I,\sP_J]= -\d_{IJ} \sJ,\\
&[\sM_{AB},\sQ_a] = \frac12 (\s_{AB})_{ab} \sQ_b,\quad [\sM_{AB},\sQ_\add]= \frac12 (\s_{AB})_{\add\bdd} \sQ_\bdd,\\
& [\sM_A,\sQ_a] = -\frac12 (\s_A)_{a\bdd} \sQ_\bdd,\quad [\sM_A,\sQ_\add]= -\frac12 (\s_A)_{\add b} \sQ_b,\\
&[\sM_{IJ},\sQ_a] = \frac12 (\s_{IJ})_{ab} \sQ_b,\quad [\sM_{IJ},\sQ_\add]= \frac12 (\s_{IJ})_{\add\bdd} \sQ_\bdd,\\
&[\sM_I,\sQ_a] = \frac12 (\s_I)_{a\bdd} \sQ_\bdd,\quad [\sM_I,\sQ_\add]= -\frac12 (\s_I)_{\add b} \sQ_b,\\
&[\sM_{AB},\bar\sQ_a] = \frac12 (\s_{AB})_{ab} \bar\sQ_b,\quad
[\sM_{AB},\bar\sQ_\add]= \frac12 (\s_{AB})_{\add\bdd} \bar\sQ_\bdd,\\
&[\sM_A,\bar\sQ_a] = -\frac12 (\s_A)_{a\bdd} \bar\sQ_\bdd,\quad
[\sM_A,\bar\sQ_\add]= -\frac12 (\s_A)_{\add b} \bar\sQ_b,\\
&[\sM_{IJ},\bar\sQ_a] = \frac12 (\s_{IJ})_{ab} \bar\sQ_b,\quad [\sM_{IJ},\bar\sQ_\add]= \frac12 (\s_{IJ})_{\add\bdd} \bar\sQ_\bdd,\\
&[\sM_I,\bar\sQ_a] = \frac12 (\s_I)_{a\bdd} \bar\sQ_\bdd,\quad
[\sM_I,\bar\sQ_\add]= - \frac12 (\s_I)_{\add b} \bar\sQ_b,\\
&[\sT,\sQ_a] = -[ \sJ , \sQ_a ] = \frac{\im}{2R} \Pi_{ab}\bar\sQ_b,\quad
[\sT,\sQ_\add]= [\sJ,\sQ_\add] = \frac{\im}{2R} \Pi_{\add\bdd}\bar\sQ_\bdd ,\\
&[\sP_A,\sQ_a] = -\frac{\im}{2R} (\s_A)_{a\bdd} \Pi_{\bdd\cdd}\bar\sQ_\cdd,\quad
[\sP_A,\sQ_\add] = -\frac{\im}{2R} (\s_A)_{\add b} \Pi_{bc}\bar\sQ_c,\\
&[\sP_I,\sQ_a] = -\frac{\im}{2R} (\s_I)_{a\bdd} \Pi_{\bdd\cdd}\bar\sQ_\cdd,\quad
[\sP_I,\sQ_\add]= -\frac{\im}{2R} (\s_I)_{\add b} \Pi_{bc}\bar\sQ_c,\\
&[\sT,\bar\sQ_a] =-[\sJ,\bar\sQ_a]= -\frac{\im}{2R} \Pi_{ab} \sQ_b,\quad
[\sT,\bar\sQ_\add]=[\sJ,\bar\sQ_\add]= -\frac{\im}{2R} \Pi_{\add\bdd} \sQ_\bdd,\\
&[\sP_A,\bar\sQ_a] = \frac{\im}{2R} (\s_A)_{a\bdd} \Pi_{\bdd\cdd} \sQ_\cdd ,\quad
[\sP_A,\bar\sQ_\add]= \frac{\im}{2R} (\s_A)_{\add b} \Pi_{bc} \sQ_c,\\
&[\sP_I,\bar\sQ_a] = \frac{\im}{2R} (\s_I)_{a\bdd} \Pi_{\bdd\cdd} \sQ_\cdd,\quad
[\sP_I,\bar\sQ_\add]= \frac{\im}{2R} (\s_I)_{\add b} \Pi_{bc} \sQ_c,\\
&\{ \sQ_a , \sQ_b \} = \d_{ab} ( \sT + \sJ ),\quad
\{ \sQ_a , \sQ_\bdd \} = \s^A_{a\bdd} \sP_A + \s^I_{a\bdd} \sP_I,\quad
\{ \sQ_\add , \sQ_\bdd \} = \d_{\add\bdd} ( \sT - \sJ ),\\
&\{ \bar\sQ_a , \bar\sQ_b \} = \d_{ab} ( \sT + \sJ ),\quad
\{ \bar\sQ_a , \bar\sQ_\bdd \} = \s^A_{a\bdd} \sP_A + \s^I_{a\bdd} \sP_I,\quad
\{ \bar\sQ_\add , \bar\sQ_\bdd \} = \d_{\add\bdd} ( \sT - \sJ ),\\
&\{ \sQ_a , \bar\sQ_b \}  = -\frac{\im}{2R} \left( (\s^{AB})_{ac} \Pi_{cb} \sM_{AB} - (\s^{IJ})_{ac} \Pi_{cb} \sM_{IJ} \right) ,\\
&\{ \sQ_a , \bar\sQ_\bdd \} = -\frac{\im}{R}\left( (\s^A)_{a\cdd} \Pi_{\cdd\bdd} \sM_A - (\s^I)_{a\cdd} \Pi_{\cdd\bdd} \sM_I \right) ,\\
&\{ \sQ_\add , \bar\sQ_\bdd \} = -\frac{\im}{2R} \left( (\s^{AB})_{\add\cdd} \Pi_{\cdd\bdd} \sM_{AB} - (\s^{IJ})_{\add\cdd} \Pi_{\cdd\bdd} \sM_{IJ} \right),\\
&\{ \sQ_\add , \bar\sQ_b \} = -\frac{\im}{R}\left( (\s^A)_{\add c} \Pi_{cb} \sM_A + (\s^I)_{\add c} \Pi_{cb} \sM_I \right) .
\end{align}

\section{Open and flat}
\label{opfl}
In this appendix we briefly review the construction of vertex operators in a light-cone gauge in  a flat background \cite{Jusinskas:2014vqa,Berkovits:2014bra}. We first start with an open pure spinor string in a flat background. The BRST gauge symmetry helps to fix the component $A_\ad$ of $A_\a$ to zero. We use the frame in which the only non-vanishing component of the momentum is $k^+$, that is, the dependence on the space-time coordinates of all superfields is $e^{-2\im k^+ X^-}$. The equations coming form $QU=0$, where $U=\l^a A_a$, are
\begin{align}
D_\ad  A_b=\s^i_{b\ad} A_i,\quad D_{(a}A_{b)}=\d_{ab} A_+ ,
\label{set1}
\end{align}
where $D_a=\p_a$ and $D_\ad=\p_\ad-ik^+\t^\ad$. A fermionic superfield $W^\a$ can be defined such that the following equations are satisfied
\begin{align}
D_a A_+=0,\quad D_a A_i=\s^i_{a\bd} W_{\bd},\quad D_\ad A_+=-2W_\ad,\quad D_\ad A_i=\s^i_{b\ad}W_b,\quad W_a=-\im k^+ A_a .
\label{set2}
\end{align}
The fermionic derivative of $W$ is related to $F=dA$, that is
\begin{align}
D_a W_b = -\frac{\im}{2}k^+\d_{ab}A_+,\quad D_\ad W_b=-\im k^+\s^i_{b\ad}A_i,\quad D_a W_\bd =0,\quad D_\ad W_\bd=\frac{\im}{2}k^+ \d_{\ad\bd}A_+ .
\label{set3}
\end{align}
Because the last equation in (\ref{set2}), the second equation in (\ref{set3}) is trivially satisfied. Applying $D_\bd$ to the last equation in (\ref{set3}) and using the third equation in (\ref{set2}) one obtains that $W_\ad=0$ and, consequently, $A_+=0$. From now on, the only physical superfields are $A_a=\frac{\im}{k^+}W_a$ and $A_i$. Note that they depend on $(\t^\ad,X^-)$ and satisfy the equations
\begin{align}
D_\ad A_b = \s^i_{b\ad} A_i,\quad D_\ad A_i = -\im k^+ \s^i_{b\ad} A_b .
\label{eq}
\end{align}
The integrated vertex operator is
\begin{align}
\oint (\p\t^a-\im k^+ d_a)A_a + (\Pi^i-2\im k^+  N^{-i})A_i ,
\label{Vopen}
\end{align}
where $N^{-i}=\frac12\s^i_{a\bd} \l^{\bd} \o_a$.

Consider the closed string case. Using the BRST gauge symmetry, the unintegrated vertex operator is $U=\l^a\lb^{\bar b} A_{a\bar b}$. The equations from $QU=0$ imply that $A_{a\bar b}$ is function of $(X^-,\t^\ad,\bar\t^{\dot{\bar a}})$ and it satisfies the equations
\begin{align}
D_\ad A_{a\bar b}=\s^i_{a\ad}A_{i\bar b},\quad D_\ad A_{i\bar b}=-\im k^+ \s^i_{a\ad} A_{a\bar b},\quad D_{\dot{\bar a}} A_{a\bar b}=\s^i_{\bar b\dot{\bar a}} A_{a i},\quad D_{\dot{\bar a}} A_{ai}=-\im k^+ \s^i_{\bar b\dot{\bar a}} A_{a\bar b} .
\label{QUc}
\end{align}
The integrated vertex is $\int d^2z~V$ and it satisfies $QV=\p\bar W-\bar\p W$, where $QW=\p U, Q\bar W=\bar\p U$ and are given by
\begin{align}
\nonumber
W=\lb^{\bar b}\left( (\p\t^a-\im k^+ d_a) A_{a\bar b}+(\Pi^i-2\im k^+ N^{-i})A_{i\bar b} \right),\\ \bar W= \l^a \left( (\bar\p \bar\t^{\bar b}-\im k^+ \bar d_{\bar b})A_{a\bar b}+(\bar\Pi^i-2\im k^+ \bar N^{-i})A_{ai} \right) ,
\label{Ws}
\end{align}
where
\begin{align}
N^{-i}=\frac12\s^i_{a\bd}\l^\bd\o_a,\quad \bar N^{-i}=\frac12\s^i_{\bar a\dot{\bar b}} \lb^{\dot{\bar b}}\ob_{\bar a} .
\label{defN}
\end{align}
The integrated vertex operator becomes
\begin{align}
\nonumber
V=&~\left( \p\t^a \bar\p \bar\t^{\bar b} -\im k^+ d_a\bar\p\bar\t^{\bar b} -\im k^+ \p\t^a \bar d_{\bar b} - (k^+)^2 d_a \bar d_{\bar b} \right) A_{a\bar b} \\
\nonumber&
+\left( \p\t^a \bar\Pi^i - 2\im k^+ \p\t^a \bar N^{-i} -\im k^+ d_a \bar \Pi^i -2(k^+)^2 d_a\bar N^{-i} \right) A_{ai}\\
\nonumber&
-\left( \Pi^i \bar\p \bar\t^{\bar a} - 2\im k^+ N^{-i} \bar\p \bar\t^{\bar a} - \im k^+ \Pi^i \bar d_{\bar a} - 2(k^+)^2 N^{-i}\bar d_{\bar a} \right) A_{i\bar a} \\
&+\left( \Pi^i \bar\Pi^j -2\im k^+ \Pi^j \bar N^{-i} -2\im k^+N^{-j} \bar\Pi^i + 8 (k^+)^2 N^{-i}\bar N^{-j} \right) A_{ij} ,
\label{ivoFLAT}
\end{align}
where
\begin{align}
A_{ij}=\frac{1}{64} \s_i^{a\bd} \s_j^{\bar a\dot{\bar b}} D_\bd
  D_{\dot{\bar b}} A_{a\bar a} .
\end{align}

{
\bibliographystyle{abe}
\bibliography{mybib}{}
}
\end{document}